\documentclass[aps,twocolumn,showpacs,pra,superscriptaddress,floatfix,longbibliography]{revtex4-1}

\usepackage{amsmath,amsthm,amsfonts,amssymb,times,bbm,graphicx,enumitem}
\usepackage{color,tabularx,epsfig,wasysym,dcolumn,bm,epstopdf,subfigure,psfrag}

\newtheorem{thm}{Theorem}[section]

\newcommand{\bd}[1]{\mbox{$\boldsymbol #1$}}
\newcommand{\bra}[1]{\mbox{$\langle #1 |$}}
\newcommand{\ket}[1]{\mbox{$| #1 \rangle$}}

\newcommand{\N}{{\mathbb{N}}}
\newcommand{\R}{{\mathbb{R}}}
\newcommand{\C}{{\mathbb{C}}}
\newcommand{\F}{{\ket{\Psi^{(f)}}}}

\begin{document}

\title{Pinning of fermionic occupation numbers: Higher spatial dimensions and spin}

\author{Felix Tennie}
\affiliation{Clarendon Laboratory, University of Oxford, Parks Road, Oxford OX1 3PU, United Kingdom}
\author{Vlatko Vedral}
\affiliation{Clarendon Laboratory, University of Oxford, Parks Road, Oxford OX1 3PU, United Kingdom}
\affiliation{Centre for Quantum Technologies, National University of Singapore, 3 Science Drive 2, Singapore 117543}
\author{Christian Schilling}
\email{christian.schilling@physics.ox.ac.uk}
\affiliation{Clarendon Laboratory, University of Oxford, Parks Road, Oxford OX1 3PU, United Kingdom}

\date{\today}

\begin{abstract}
The role of the generalized Pauli constraints (GPCs) in higher spatial dimensions and by incorporating spin degrees of freedom is systematically explored for a system of interacting fermions confined by a harmonic trap. Physical relevance of the GPCs is confirmed by analytical means for the ground state in the regime of weak couplings by finding its vector of natural occupation numbers close to the boundary of the allowed region. Such quasipinning is found to become weaker in the intermediate and strong coupling regime. The study of crossovers between different spatial dimensions by detuning the harmonic trap frequencies suggests that quasipinning is essentially an effect for systems with reduced spatial dimensionality. In addition, we find that quasipinning becomes stronger by increasing the degree of spin-polarization. Consequently, the number of states available around the Fermi level plays a key role for the occurrence of quasipinning. This suggests that quasipinning emerges from the conflict between energy minimization and fermionic exchange symmetry.
\end{abstract}


\maketitle

\section{Introduction}\label{sec:intro}

\begin{figure}[]
\centering
\includegraphics[width=8.5cm]{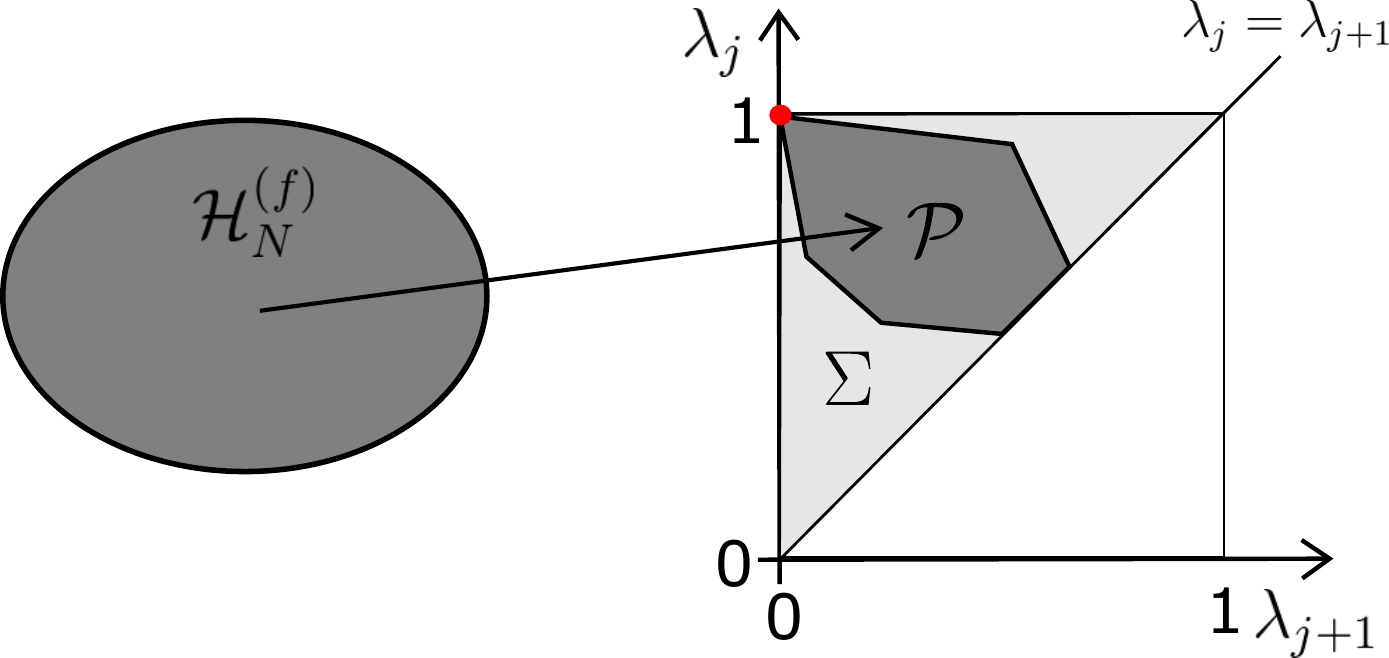}
\caption{Illustration of the mapping of $N$-fermion quantum states (left) onto their vectors $\vec{\lambda}$ (right) of decreasingly-ordered natural occupation numbers, giving rise to a polytope $\mathcal{P}$ (dark grey). $\mathcal{P}$ is a proper subset of the Pauli-simplex $\Sigma$ (grey) defined by $1\geq \lambda_1\geq \ldots\geq \lambda_d\geq 0$.}
\label{fig:polytope4}
\end{figure}

Pauli's exclusion principle (PEP) has been a strong guiding tool for the understanding and description of numerous many-fermion systems. It however does not entirely resemble the exchange antisymmetry of the $N$-fermion wave function. In a number of works \cite{Borl1972,Kly2,Kly3,Altun,Rus2} that property was found to impose greater restrictions on the \emph{1-particle reduced density operator} ($1$-RDO):
The vector $\vec{\lambda}\equiv (\lambda_i)_{i=1}^d$ of its decreasingly-ordered eigenvalues $\lambda_i$, the so-called \emph{natural occupation numbers} (NONs), is confined to a polytope within the `Pauli simplex' $\Sigma$ defined by Pauli's exclusion principle $1\geq \lambda_1\geq \ldots \geq \lambda_d\geq 0$ (see Figure \ref{fig:polytope4}). The corresponding  constraints take the form of linear inequalities:
\begin{equation}\label{eq:gpc}
D_j(\vec{\lambda}) \equiv \kappa_j^{(0)}+\vec{\kappa}_j \cdot  \vec{\lambda}\,\geq \,0\,,\quad j=1,2,\ldots,r_{N,d},
\end{equation}
and are commonly referred to as \emph{generalized Pauli constraints} (GPCs).

Exploring the physical significance of the GPCs remains to be a challenge that has recently seen a growing interest among physicists and quantum chemists \cite{Kly1,CS2013,Kly5,BenavLiQuasi,Mazz14,CSthesis,Mazzopen,CSQuasipinning,BenavQuasi2,RDMFT,Alex,MazzCopy,CS2015Hubbard,CSQ,Benavbirthday,CS2016a,Mazz16}. Especially the phenomenon where the $\lambda$-vector lies close or on the boundary \footnote{Here and in the following, `boundary' refers only to that part of the mathematical boundary $\partial\mathcal{P}_{N,d}$ of the polytope $\mathcal{P}_{N,d}$ which corresponds to saturation of a GPC. The part of $\partial\mathcal{P}_{N,d}$ corresponding to saturation of an ordering constraint, $\lambda_i-\lambda_{i+1}\geq 0$ or $\lambda_d\geq 0$ will not play any role.} of the polytope, also known as \emph{quasipinning} \cite{CS2013,CSthesis,CSQuasipinning} and \emph{pinning} \cite{Kly1}, respectively, has stimulated much research since it implies a number of remarkable properties for the corresponding $N$-fermion quantum system \cite{CSQMath12}. The most striking implication is the resulting structural simplification of the $N$-fermion quantum state. Therefore, studying and understanding (quasi)pinning may provide further insights into concepts as entanglement \cite{entangl36,entangl39,Chen2014a,Chen2014b,entanglCC} and correlations \cite{AlexCmeas,AlexNonfreeness1,AlexNonfreeness2,CIforBD} in few-fermion quantum systems as being recently explored from a quite conceptual and often particularly quantum information theoretical viewpoint.

This paper aims to explore in great detail the scope of (quasi)pinning and to recognize and describe its origin. In Ref.~\cite{CS2016a}, tools and methods required for a sound and conclusive (quasi)pinning analysis have been introduced and applied to the instructive model system of Harmonium in one spatial dimension. In the present paper, we are going to extend these investigations to systems of higher spatial dimensions and to spinful particles. Furthermore, dimensional crossovers will be studied by considering detunings of the external trapping frequencies. Alongside the results in Ref.~\cite{CS2016a}, this will suggest that the structure of the active space around the Fermi level plays a key role in the occurrence of quasipinning and will particularly show that quasipinning is becoming stronger by reducing the spatial dimensionality.

The paper is organized as follows: In Section \ref{sec:model} the model of Harmonium in multiple spatial dimensions will be introduced and its ground state will be determined. The definition, form and properties of the $1$-particle reduced density operator and related NONs will be discussed in Section \ref{sec:1RDM}. This is followed by Section
\ref{sec:1dquasi} where quasipinning is explored for a quasi-one-dimensional system. The extension to proper two and three-dimensional systems will be provided in Section \ref{sec:Harmonium23dNoSpin} by a thorough analysis. Results of (quasi)pinning analyses of different dimensional crossovers will be shown as well. In the sixth section, spin degrees of freedom will be incorporated by considering non-fully polarized fermions and magnetic crossovers between different spin states. Section \ref{sec:concl} summarizes and discusses our findings.

\section{Model and its ground state}\label{sec:model}
In this section we generalize the model of Ref.~\cite{CS2016a} to higher spatial dimensions. We discuss in detail the structure of its ground state and define the relevant coupling parameters.

The system we are considering consists of $N$ identical (yet spinless) particles of mass $m$ that are confined in an $n$-dimensional (not necessarily isotropic) harmonic trapping potential characterized by its trapping frequencies $\left\{ \omega^{(\alpha)}\right\}_{ \alpha = 1}^n$. In addition to the external potential, a harmonic particle-particle interaction of strength $K$ will be taken into account. Consequently, the Hamiltonian reads
\begin{align}\label{eq:ham}
H_N = \sum_{i=1}^{N} \left(\frac{\vec{p}_i^{\,2}}{2 m} +\frac{m}{2} {\vec{x}_i}^{\,t}\Omega {\vec{x}_i}\right) + \frac{K}{2} \sum_{1\leq i<j\leq N} (\vec{x}_i-\vec{x}_j)^2,
\end{align}
where $\vec{p}_i = (p_i^{(\alpha)})_{\alpha = 1}^n$ and $\vec{x}_i = (x_i^{(\alpha)})_{\alpha = 1}^n$ represent the momentum and position operators of the $i$-th particle, respectively, and $\Omega\equiv\mbox{diag}(\omega_1^2,\ldots,\omega_n^2)$. In the present paper, the \emph{N-fermion} system  associated with this Hamiltonian is referred to as Harmonium.

The Harmonium model arises as an effective model in the harmonic approximation applied to systems of harmonically confined interacting particles \cite{koscikapprox}. For instance, this approximation works fairly well in the description of quantum dots since the Coulomb interaction between the electrons is screened (see Ref.~\cite{QuantumDot}). Furthermore, Harmonium was used to explore the emergence of shell structures in atoms (see e.g.~Ref.~\cite{HarmShells}) and nuclei (see e.g.~Ref.~\cite{ZinnerNucl}). Also for ultracold gases in a trap it plays a role since there the two-body interactions can be tuned \cite{ultracoldBook1,BlochReview,Feshbach}. In particular, it is possible to change the interactions from repulsive to attractive \cite{Jochim2species}.

Harmonium encompasses an important advantage. In order to analytically investigate the occurrence of quasipinning, a system must permit the \textit{analytical} execution of all of the following steps:
\begin{enumerate}[label=(\roman*)]
\item computation of the many-body eigenstates in a system of \emph{interacting fermions} (see e.g.~Refs.\cite{harmOsc2012,CS2013,CS2013NO,ZinnerNbody})
\item tracing out $N-1$ particles (i.e.~calculating analytically $(N-1) n$ integrals)
\item diagonalizing the $1$-RDO, either analytically in the regime of weak couplings or exact numerically for medium and strong couplings.
\end{enumerate}
While all three steps are possible for Harmonium, none of them is feasible for most continuous physical models.

\emph{A priori}, the Hamiltonian \eqref{eq:ham} acts as an operator on the $N$-particle Hilbert space $\mathcal{H}^{(N)} = \bigotimes_{i=1}^{N}\mathcal{H}$,
where the $1$-particle Hilbert space $\mathcal{H}$ is given by $\mathcal{H} = L_2(\mathbb{R}^n)$. When incorporating spin degrees of freedom in Section \ref{sec:spin}, this will be modified to $\mathcal{H} = L_2(\mathbb{R}^n)\otimes\mathbb{C}^2$. Any permutation of particles leaves $H_N$ invariant. In particular, this allows us to treat the $N$ particles as indistinguishable fermions and thus restrict the Hamiltonian \eqref{eq:ham} to the subspace
\begin{equation}\label{eq:hspaces}
\mathcal{H}_N^{(f)} \equiv \wedge^N\left[\mathcal{H}\right] = \mathcal{A}_N\mathcal{H}_N \lneq \mathcal{H}_N\equiv \mathcal{H}^{\otimes^N}
\end{equation}
of antisymmetric states. Here, $\mathcal{A}_N$ represents the antisymmetrising operator. In order to derive the set of fermionic eigenstates of the Hamiltonian \eqref{eq:ham} we therefore initially may derive the set of all $N$-particle eigenstates followed by a projection onto $\mathcal{A}_N\mathcal{H}_N$.

In the following we present and describe the fermionic ground state and provide a concise derivation in Appendix \ref{app:gs}. It turns out to be instructive to first discuss the case of zero interaction, $K=0$. Clearly, in that case the corresponding time-independent Sch\"odinger equation for the Hamiltonian (\ref{eq:ham}) effectively simplifies to a $1$-fermion eigenvalue problem,
\begin{equation}\label{eq:ham1fermion}
\left(\frac{\vec{p}^{\,2}}{2m} +\frac{1}{2}m\vec{x}^t\Omega\vec{x}\right)\chi=\varepsilon\, \chi\,.
\end{equation}
The solutions of Eq.~(\ref{eq:ham1fermion}) are given by the $n$-dimensional Hermite functions denoted by $\phi_{\bd{\mu}}^{(\bd{l})}(\vec{x})\equiv \prod_{\alpha=1}^n \varphi_{\mu^{(\alpha)}}^{(l^{(\alpha)})}(x^{(\alpha)})$ with corresponding energy $\varepsilon_{\bd\mu}=\sum_{\alpha=1}^n(\mu^{(\alpha)}+\frac{1}{2})\hbar\omega^{(\alpha)}$. Here, $\varphi_{m}^{(l^{(\alpha)})}$ denotes the $m$-th Hermite function in one dimension with natural length scale $l^{(\alpha)} \equiv \sqrt{\frac{\hbar}{m\omega^{(\alpha)}}}$ , $\bd\mu\equiv (\mu^{(\alpha)})_{\alpha=1}^n$ and $\bd{l}\equiv (l^{(\alpha)})_{\alpha=1}^n$.
The fermionic eigenstates of the Hamiltonian (\ref{eq:ham}) for zero interaction are then given by the `configuration states', i.e.~by Slater determinants obtained by distributing the $N$ particles in $N$ different states $\phi_{\bd{\mu}}^{(\bd{l})}$. In particular, the corresponding ground state is then given by filling the $N$ $1$-particle states with lowest $1$-particle energies $\varepsilon_{\bd \mu}$ according to Pauli's exclusion principle.

In general, for interacting fermions one cannot expect that the structure of the energy eigenstates can be elegantly described by exploiting the elementary and convenient $1$-fermion picture. Yet, a bit surprisingly, this is still possible at least for the ground state of Harmonium \footnote{The case of an isotropic trap, $\omega^{(1)}=\ldots=\omega^{(n)}$, was already discussed in Ref.~\cite{harmOsc2012} but no constructive proof was presented.}:
\begin{thm}\label{thm:gs}
The $N$-fermion ground state $\Psi^{(f)}$ of the Harmonium model (\ref{eq:ham}) is given by
\begin{equation}\label{eq:gs}
\Psi^{(f)}(\vec{x}_1,\ldots,\vec{x}_N)=\mathcal{N}\,\left|
\begin{array}{ccc}
\phi_{\bd{\mu}_1}^{(\bd{\tilde{\bd{l}}})}(\vec{x}_1)&\cdots&\phi_{\bd{\mu}_1}^{(\bd{\tilde{\bd{l}}})}(\vec{x}_N) \\
\vdots&&\vdots\\
\phi_{\bd{\mu}_N}^{(\bd{\tilde{\bd{l}}})}(\vec{x}_1)&\cdots&\phi_{\bd{\mu}_N}^{(\bd{\tilde{\bd{l}}})}(\vec{x}_N) \end{array}
\right|\times e^{\vec{X}^t\bd{B}\vec{X}}\,,
\end{equation}
with $\vec{X}\equiv \frac{1}{N}(\vec{x}_1+\ldots \vec{x}_N)$ the center of mass vector, $\bd{B}\equiv \mbox{diag}(B^{(1)},\ldots,B^{(n)})$, $B^{(\alpha)}\equiv  \frac{N}{2}\left(\frac{1}{(\tilde{l}^{(\alpha)})^2}
-\frac{1}{(l^{(\alpha)})^2}\right)$, $\tilde{\omega}^{(\alpha)}\equiv \sqrt{(\omega^{(\alpha)})^2+\frac{NK}{m}}$,
$\tilde{l}^{(\alpha)} \equiv \sqrt{\frac{\hbar}{m \tilde{\omega}^{(\alpha)}}}$ and $\mathcal{N}$ is a normalization constant.
The quantum number vectors $\bd{\mu}_1,\ldots,\bd{\mu}_N$ in \eqref{eq:gs} are chosen such that the following energy function
\begin{equation}\label{eq:energy}
E_{\bd{\mu}_1,\ldots,\bd{\mu}_N}\equiv \sum_{i=1}^N \tilde{\varepsilon}_{\bd{\mu}_i}
\end{equation}
is minimal, yet respecting Pauli's exclusion principle \footnote{Strictly speaking, this is not exactly the Pauli exclusion principle since it is not applied to the total quantum state $\Psi^{(f)}$  but just to one of its two factors. In a similar way, one should understand the term `distributing $N$ fermions in $N$ shells' more as an analogy.} (i.e.~all $\bd{\mu}_i$ are different)
and $\tilde{\varepsilon}_{\bd{\mu}_i}\equiv \sum_{\alpha=1}^n(\mu^{(\alpha)}+\frac{1}{2})\,\hbar\tilde{\omega}^{(\alpha)}$.
\end{thm}

Since the proof of Theorem \ref{thm:gs} is less trivial we present it in Appendix \ref{app:gs}. As a caveat, we would like to stress that the excited states are \emph{not} given by filling `boxes' of higher energy and then multiplying the corresponding Slater determinant by the exponential factor as in Eq.~(\ref{eq:gs}). Indeed, their structure is more complicated and the single Slater determinant in Eq.~(\ref{eq:gs}) would need to be replaced by a linear combination of several Slater determinants, expressing the additional correlations in the system.

\begin{figure}[]
\centering
\includegraphics[width=8.0cm]{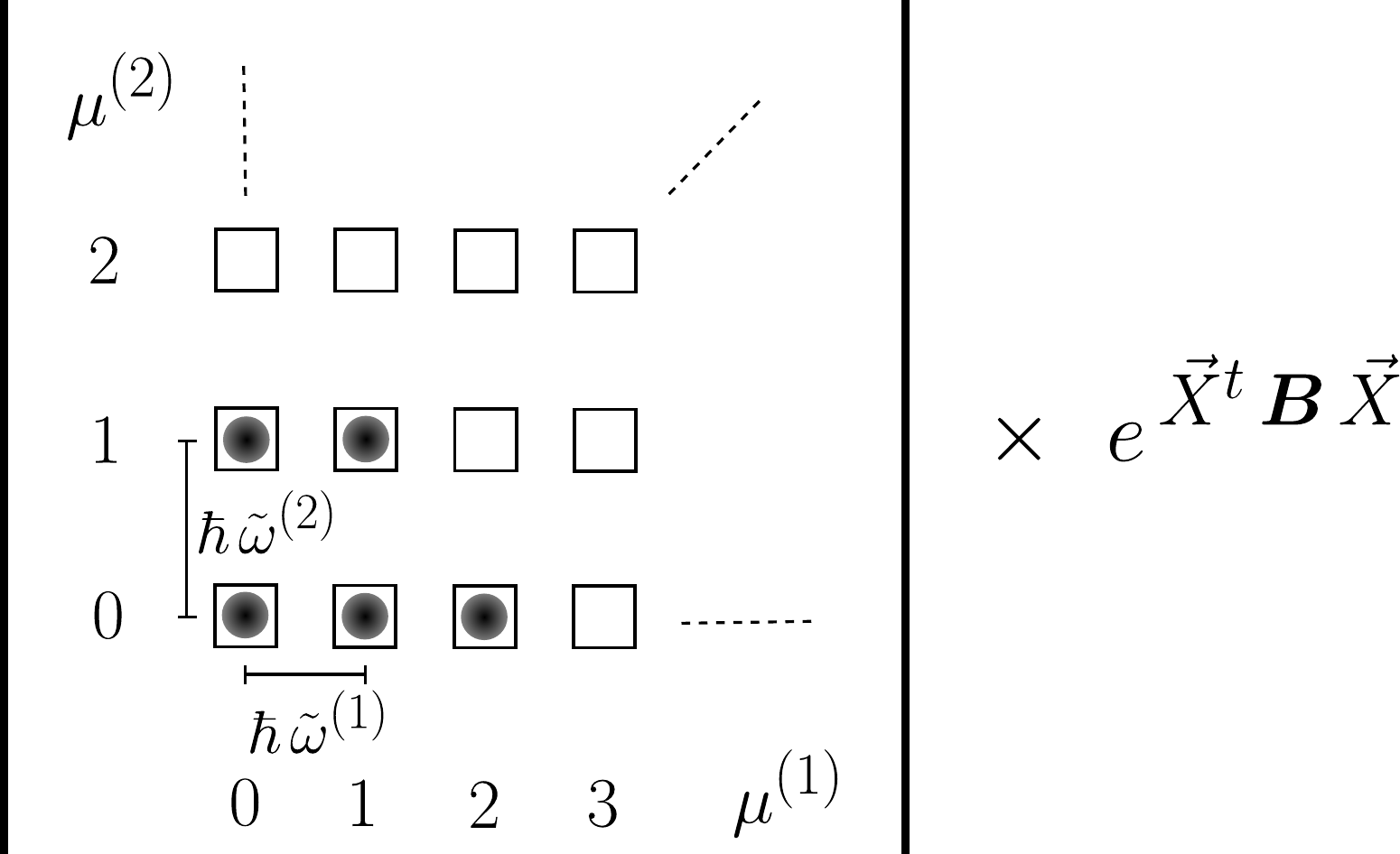}
\caption{Graphical illustration of the fermionic ground state $\ket{\Psi^{(f)}}$ for the exemplary case of $n=2$ spatial dimension and $N=5$ fermions. In general, $\ket{\Psi^{(f)}}$ is given by a single Slater determinant, obtained by filling the $N$ `boxes' with lowest energy respecting Pauli's exclusion principle (left), multiplied by a correlation term $e^{\vec{X}^tB\vec{X}}$ (right) for the center of mass $\vec{X}\equiv \frac{1}{N}(\vec{x_1}+\ldots \vec{x}_N)$. Each box, labeled by $\vec{\mu}=(\mu_1,\ldots,\mu_n)$ describes a $1$-particle orbital given by a harmonic oscillator state in $n$-dimensions with corresponding frequencies $\tilde{\omega}^{(\alpha)}$, $\alpha=1,\ldots,n$.}
\label{fig:box}
\end{figure}

The ground state as stated in Theorem \ref{thm:gs} is graphically illustrated in Figure \ref{fig:box}. It is given by a Slater determinant, obtained by successively distributing fermions in the `boxes' (energy levels) $\phi_{\bd \mu}^{(\tilde {\bd l})}$ with lowest energy $\tilde{\varepsilon}_{\bd{\mu}}$, respecting Pauli's exclusion principle, multiplied by the term $e^{\vec{X}^t\bd{B}\vec{X}}$. This latter term contains the physical correlations, in contrast to the Slater determinant which contains only exchange correlations.
For the example in Figure \ref{fig:box} of $N=5$ and $n=2$ the occupied `boxes' are given by the `configuration' $\mathcal{C}=\{\{0,0\},\{1,0\},\{0,1\},\{2,0\},\{1,1\}\}$.

Consequently, the ground state for finite interaction has some similarity to the ground state for zero interaction. They differ by the correlation term $e^{\vec{X}^t\bd{B}\vec{X}}$ and a change of the natural length scales $\tilde{l}^{(\alpha)}$ of the Hermite functions $\Phi_{\bd \mu}^{(\bd{\tilde{\bd l}})}$. Furthermore, if the coupling constant $K$ (or some $\omega^{(\alpha)}$) changes, the structure of the ground state can change at some `critical' values: For this, notice that the energy ratios $\tilde{\varepsilon}_{\bd \mu}/\tilde{\varepsilon}_{\bd{\mu'}}$, i.e.~the ratios of the distances between the boxes in different directions $\alpha=1,\ldots,n$ in Figure \ref{fig:box}, depends on the coupling strength. By changing these ratios,
the $N$-th lowest and $(N+1)$-th lowest energies $\bd{\tilde{\varepsilon}_{\bd \mu}}$ may cross and the $N$ fermions change their configuration $\{\bd{\mu}_1,\ldots,\bd{\mu}_N\}$ to minimize the total energy (\ref{eq:energy}). Such crossings particularly occur when one of the harmonic trap frequencies $\omega^{(\alpha)}$ is increased to very large values implying that all degrees of freedom in the corresponding $\alpha$-dimension are frozen. For instance, in the example presented in Figure \ref{fig:box} increasing $\omega^{(2)}$ will initially change the ground state configuration to $\{\{0,0\},\{1,0\},\{2,0\},\{0,1\},\{3,0\}\}$ and eventually to the quasi-one-dimensional case $\{\{0,0\},\{1,0\},\{2,0\},\{3,0\},\{4,0\}\}$.

As it was illustrated in Ref.~\cite{CS2016a} it is instructive to compare the fermionic ground state $\Psi^{(f)}$ to the bosonic ground state $\Psi^{(b)}$ of the Hamiltonian (\ref{eq:ham}). The bosonic ground state can be described by the `box picture' as well: $\Psi^{(b)}$ is given by distributing all $N$ bosons in the lowest energy `box' (defined by $\bd{\mu}\equiv(0,\ldots,0)\equiv \emph{\bd{0}}$) and multiplying the corresponding product state by the same exponential $e^{\vec{X}^t\bd{B}\vec{X}}$ as for fermions (c.f.~Eq.~\ref{eq:gs}). Consequently, we get (see also Appendix \ref{app:gs})
\begin{eqnarray}\label{eq:gsb}
\Psi^{(b)}(\vec{x}_1,\ldots,\vec{x}_N) &\sim& \left(\prod_{i=1}^N\,\phi_{\emph{\bd{0}}}^{(\bd{\tilde{\bd{l}}})}(\vec{x}_i)\right)\times e^{\vec{X}^t\bd{B}\vec{X}}\nonumber \\
&\sim& \left(\prod_{i=1}^N\,e^{-\vec{x}_i\,\bd{A}\,\vec{x}_i}\right)\times e^{\vec{X}^t\bd{B}\vec{X}}
\end{eqnarray}
where $\bd{A}\equiv \mbox{diag}( A^{(1)},\ldots,A^{(n)})$ and $A^{(\alpha)}\equiv \frac{1}{2 (\tilde{l}^{(\alpha)})^2}$.

It is also worth discussing the symmetries of the $N$-Harmonium ground state (\ref{eq:gs}).
For all parameters $K$, $\{\omega^{(\alpha)}\}_{\alpha=1}^n$ the Hamiltonian (\ref{eq:ham}) is invariant under simultaneous spatial reflections $P^{(\alpha)}:x_i^{(\alpha)}\rightarrow -x_i^{(\alpha)}$, $i=1,\ldots,N$, of the $\alpha$-coordinate of various fermions, i.e.
\begin{equation}
[\hat{H},U(P^{(\alpha)})^{\otimes^N}]=0\,,
\end{equation}
where $U$ is its unitary representation on the $1$-particle Hilbert space $\mathcal{H}$. Consequently, in particular the ground state (\ref{eq:gs}) inherits the corresponding symmetry, $U(P^{(\alpha)})^{\otimes^N}\Psi^{(f)}=\pm \Psi^{(f)}$.

Furthermore, if two trap frequencies are equal, $\omega^{(\alpha)}=\omega^{(\alpha')}$ the Hamiltonian is in addition also invariant under simultaneous spatial rotation  $R^{(\alpha,\alpha')}:(x_i^{(\alpha)},x_i^{(\alpha')})^t \rightarrow R \,(x_i^{(\alpha)},x_i^{(\alpha')})^t$, $R \in SO(2)$, of the $\alpha$- and $\alpha'$-coordinates of all $N$ fermions. This extends in an elementary way to the case of more than two identical trap frequencies. For the case of isotropic traps, shell structures similar to those in atoms emerge \cite{HarmShells}.

\section{$1$-particle reduced density operator and natural occupation numbers}\label{sec:1RDM}
In this section, we discuss symmetries of $1$-RDOs in general and present specific properties of the $1$-RDO of the $N$-Harmonium ground state in particular.

\subsection{General considerations}\label{sec:1RDMgeneral}
For pure $N$-fermion quantum states $\ket{\Psi^{(f)}}$ the $1$-RDO is defined by
\begin{equation}\label{eq:1RDO}
\rho^{(f)} = N\,\mbox{tr}_{N-1}[\ket{{\Psi^{(f)}}}\bra{{\Psi^{(f)}}}]
\end{equation}
and is consequently normalized to the particle number $N$, i.e.~$\mbox{tr}[\rho^{(f)}] \: =\: N$. Due to the fermionic exchange symmetry the choice of the $N-1$ fermions to be traced out does not matter.

It is worth noting that $1$-RDOs inherit $1$-particle symmetries of the corresponding $N$-fermion state \cite{Dav}. To explain this, consider a $1$-particle symmetry of the $N$-fermion quantum state $\ket{\Psi^{(f)}}$ generated by the unitary operator $G=g^{\otimes^N}$ with $g$ acting on the $1$-particle Hilbert space $\mathcal{H}$,
\begin{equation}\label{eq:symN}
  G\ket{\Psi^{(f)}} = e^{i \zeta}\ket{\Psi^{(f)}}\,,
\end{equation}
with $\zeta \in \R$. Then, due to elementary properties of partial traces, the $1$-RDO $\rho^{(f)}$ of $\ket{\Psi^{(f)}}$ inherits this symmetry and it follows
\begin{equation}\label{eq:sym1}
[g,\rho^{(f)}] = 0\,.
\end{equation}
Eq.~(\ref{eq:sym1}) particularly implies that the $1$-RDO expressed as a matrix with respect to the $g$-symmetry-adapted states (eigenstates of $g$) is block-diagonal. For instance, for translationally invariant $N$-electron states for $1$-band lattice models this implies that the natural orbitals, the eigenstates of $\rho^{(f)}$, are given by the Bloch states multiplied by a spin state. In that case, the NONs gain a lot of physical significance \cite{CS2015Hubbard} since they are not just the eigenvalues of the $1$-RDO but the occupancies with respect to physically distinguished $1$-particle states (Bloch states).

\subsection{$1$-particle reduced density operator for the Harmonium ground state}\label{sec:1RDMspecific}
Applying the considerations of Section \ref{sec:1RDMgeneral} to the parity symmetries of $N$-Harmonium we can conclude that the $1$-RDOs of various eigenstates are block-diagonal with respect to each parity-symmetry $U(P^{(\alpha)})$, $\alpha=1,2,\ldots,n$. For instance, for the $1$-RDO $\rho^{(f)}$ of the $N$-Harmonium ground state $\F$ in two spatial dimensions we find
\begin{equation}\label{eq:1RDOparities}
\rho^{(f)} = \rho_{\boldsymbol{ee}}^{(f)}\oplus \rho_{\boldsymbol{eo}}^{(f)}\oplus \rho_{\boldsymbol{oe}}^{(f)}\oplus \rho_{\boldsymbol{oo}}^{(f)},
\end{equation}
where the indices $\boldsymbol{e}$ and $\boldsymbol{o}$ denote underlying even and odd parities for the respective dimensions $\alpha=1,2$. Decomposition structures as (\ref{eq:1RDOparities}) will help us to significantly simplify the calculation of the eigenvalues of the $1$-RDO $\rho^{(f)}$.
\\
\\
\\
Similar to the computations in Ref.~\cite{CS2016a} for the ground state of Harmonium in one spatial dimension we determine the $1$-RDO of the ground state in higher spatial dimensions by using the Hubbard-Stratonovitch relation (see Appendix \ref{app:1RDO} for more details).
Again, it is instructive to relate the fermionic  $1$-RDO $\rho^{(f)}$ to the $1$-RDO $\rho^{(b)}$ of the bosonic ground state (\ref{eq:gsb}). By comparing Eqs.~\eqref{eq:gs}, \eqref{eq:gsb} one immediately finds
\begin{equation}\label{eq:1RDMfb}
\rho^{(f)}(\vec{x},\vec{y})= F(\vec{x},\vec{y})\cdot \rho^{(b)}(\vec{x},\vec{y}) \,,
\end{equation}
where $F(\vec{x},\vec{y})$ is a multivariate polynomial and $\rho^{(b)}(\vec{x},\vec{y})$ is a Gaussian.

Since $\rho^{(b)}$ is a Gaussian state it can be analytically diagonalized (see for instance Ref.~\cite{CS2013NO}). In contrast to $\rho^{(b)}$, this is not possible for the fermionic $1$-RDO $\rho^{(f)}$. Yet, similar to Ref.~\cite{CS2016a}, we can diagonalize $\rho^{(f)}$ by numerical means for fixed couplings.
We introduce dimensionless coupling strengths for the spatial dimensions $\alpha=1,2,\ldots,n$,
\begin{equation}\label{eq:kappa}
  \kappa^{(\alpha)} \equiv \frac{N K}{m (\omega^{(\alpha)})^2} = \left(\frac{l^{(\alpha)}}{\tilde{l}^{(\alpha)}}\right)^4-1\,.
\end{equation}
Recall, that the structure of the Hamiltonian (\ref{eq:ham}) implies that the NONs of any of its eigenstates do not depend on $m (\omega^{(\alpha)})^2$ and $K$ separately, but just on their ratios.

In addition, we use again as in Refs.~\cite{CS2013,CS2016a} a perturbational approach for the regime of weak couplings. This perturbation theoretical approach can be simplified by exploiting a duality of NONs proven in Ref.~\cite{duality}: By employing the alternative coupling parameters
\begin{eqnarray}\label{eq:delta}
\delta^{(\alpha)}:=\ln\left(\frac{l^{(\alpha)}}{\tilde l^{(\alpha)}}\right) &=& \frac{1}{4}\ln\left(1+ \kappa^{(\alpha)}\right) \nonumber \\
&=&\frac{1}{4}\kappa^{(\alpha)} +\mathcal{O}((\kappa^{(\alpha)})^2)\,,
\end{eqnarray}
with $\alpha=1,2,\ldots,n$, this duality reads
\begin{equation}\label{eq:duality}
\vec{\lambda}(\ldots,\delta^{(\alpha)},\ldots)=\vec{\lambda}(\ldots,-\delta^{(\alpha)},\ldots)\,,
\end{equation}
for each $\alpha=1,2,\ldots,n$.
As a consequence, the series expansions of various NONs simplifies since it contains even orders of each $\delta^{(\alpha)}$, only.

We conclude this section by briefly recalling that the exponential factor $\rho^{(b)}$ in the expression (\ref{eq:1RDMfb}) implies an exponential decaying behavior of the NONs (for more details we refer the reader to Section IV.B in Ref.~\cite{CS2016a}). This allows us to simplify the investigations of possible (quasi)pinning by exploiting the concept of truncation described in Ref.~\cite{CS2016a}: The infinite spectrum of NONs can be truncated by skipping various NONs which are very close to $1$ or $0$, respectively, and the (quasi)pinning analysis can be performed within the remaining smaller setting. Any result on possible quasipinning found in the truncated setting translates to quasipinning of the same strength in the infinite setting up to a small truncation error (for more details see Ref.~\cite{CS2016a}).

In the following sections we explore the occurrence of (quasi)pinning in great detail for various spatial dimensions, different particle numbers and last but not least by incorporating spin degrees of freedom.

\section{Quasi-one-dimensional Harmonium}\label{sec:1dquasi}
Before investigating truly higher-dimensional systems in later sections, we are going to explore the quasipinning behaviour in quasi-one-dimensional systems. This will especially allow us to link outcomes of subsequent analyses to the results from Refs.~\cite{CS2013,CS2016a} for the strictly one-dimensional Harmonium. By quasi-one-dimensional systems we mean systems in which all but one trapping frequencies have been highly detuned. In addition, we assume those detuned frequencies $\omega^{(2)},\ldots,\omega^{(n)}$ to be identical yielding a hypercylindric trapping potential. In graphical representations of the Harmonium ground state as shown in Fig.~\ref{fig:box} this relates to a situation in which the inter-`box' spacings $\hbar \tilde{\omega}^{(\alpha)}$ have severely been increased in all but one dimension, $\tilde{\omega}^{(\alpha)}/\tilde{\omega}^{(1)}\gg 1$ for $\alpha = 2, \ldots,N$. Consequently, only energy levels along the $1$-axis will be occupied. This resembles the physical situation in which harmonic oscillators in the directions $\alpha \geq 2$ carry no excitations.

Generally, the notion of effective lower dimensionality can be captured more formally as follows. If a configuration has quantum number vectors $\{\bd{\mu}_i\}_{i=1}^{N}$ with \mbox{$\mu^{(\alpha)}_{i}=0$} for all $\alpha>n'$ and $i = 1\ldots N$, the fermionic ground state function decomposes into a product of the fermionic ground state function in the first $n'$ dimensions and the bosonic ground state function in the remaining $(n'-n)$ dimensions,
\begin{equation}\label{eq:EffectiveLowerDimensionalityOnWavefunctionLevel}
\Psi^{(f)}_n = \Psi^{(f)}_{n'}\otimes\Psi^{(b)}_{n-n'}.
\end{equation}
Such effective lower dimensionality has structural implications on the fermionic $1$-RDO as well: The polynomial $F$ in Eq.~\eqref{eq:1RDMfb} will not depend on $x^{(\beta)}_{i}, y{}^{(\beta)}_{i}$ with $i = 1\ldots N$ and  $\beta>n'$. Hence, the $1$-RDO becomes a product of a fermionic $1$-RDO in the first $n'$ dimensions and the bosonic $1$-RDO in remaining $n-n'$ dimensions,
\begin{equation}\label{eq:lowerDimStructuralImplicationOnFermionic1RDO}
\rho^{(f)}_n = \frac{1}{N}\, \rho^{(f)}_{n'}\otimes \rho^{(b)}_{n-n'}.
\end{equation}
The factor $N$ originates from the specific normalization of $1$-RDOs, $\mbox{tr}[\rho]=N$.

As a consequence of Eq.~(\ref{eq:lowerDimStructuralImplicationOnFermionic1RDO}), the spectrum of $\rho_{n}^{(f)}$ becomes the product of
the spectrum of the lower-dimensional fermionic $1$-RDO and the spectrum of the bosonic $1$-RDO,
\begin{equation}\label{eq:productSpectrum}
\vec{\lambda}^{(f)}_n = \frac{1}{N} \left(\vec{\lambda}^{(f)}_{n'}\times \vec{\lambda}^{(b)}_{n-n'}\right)^{\downarrow},
\end{equation}
where `$\downarrow$' denotes decreasing ordering.

These considerations can now be applied to quasi-one-dimensional Harmonium which is achieved for sufficiently large trapping frequency detunings, namely $\tilde{\omega}^{(\alpha)} > (N-1)\,\tilde{\omega}^{(1)}$ for $\alpha\geq 2$. The corresponding vector of ground state NONs is then given by the product of the spectrum $\vec{\lambda}_1^{(f)}$ of the one-dimensional fermionic $1$-RDO and $(n-1)$ spectra $\vec{\lambda}_1^{(b)}$ of the ($n-1)$ decoupled one-dimensional bosonic $1$-RDOs. According to Ref.~\cite{CS2016a}, the spectrum of the one-dimensional bosonic $1$-RDO with coupling $\overline{\kappa}$ is given by the NONs
\begin{equation}\label{eq:NONb}
\lambda^{(b)}_{k}(\overline{\kappa}) = N(1-q(\overline{\kappa}))\,q(\overline{\kappa})^k\,,\quad k\in \mathbb{N}_0^+\,,
\end{equation}
where
\begin{equation}\label{eq:q}
q(\overline{\kappa}) = 1 - \frac{2N}{N+\sqrt{N^2-(N-1)\left[2-(1+\overline{\kappa})^2-1/(1+\overline{\kappa})^2\right]}}\,.
\end{equation}
The corresponding coupling strengths $\kappa^{(2)}=\ldots=\kappa^{(n)}\equiv \overline{\kappa}$ for those bosonic spectra can be expressed as function of $\delta\equiv \delta^{(1)}$ and the trapping frequency ratios $\chi\equiv \omega^{(\alpha)}/\omega^{(1)}$, $\alpha\geq2$.
It is also worth noticing that with increasing detunings $\chi$, the bosonic spectra approach more and more $(N,0,0,\ldots)$.

We apply these general ideas to the exemplary case $N=3$ and $n=2$. The spectrum $\vec{\lambda}_1^{(f)}$ was determined in Ref.~\cite{CS2013} and is also listed in Appendix \ref{app:NONsPRL}. Then, we determine the NONs of the corresponding $3$-Harmonium ground state via Eq.~(\ref{eq:productSpectrum}).
The corresponding quasipinning analysis is slightly involved. The main reason for this is that by varying the parameters $\delta$ and $\chi$ the hierarchy of NONs in Eq.~(\ref{eq:productSpectrum}) does change, which changes the quasipinning behavior as well \cite{CS2015Hubbard}. Even within regimes of non-crossing NONs, providing a closed analytical expression for the minimal distance $D_{min}$
of $\vec{\lambda}$ to the polytope boundary is often not possible since the GPC which is most saturated may change while changing $\delta$ and $\chi$.

This then requires to split the parameter space into separate regimes where the most relevant NONs do not cross. We consider in the following only the regime which includes particularly the limit $\chi\rightarrow\infty$.
This regime turns out to be defined by
\begin{equation}\label{eq:regimecondition}
\chi>\chi_{crit}(\delta)\equiv \sqrt[4]{\frac{243}{40}}\,\frac{1}{\delta^{\frac{3}{2}}}\,,
\end{equation}
where $\delta\ll1$.
A thorough and quite lengthy quasipinning analysis is performed by exploiting the concept of truncation (c.f.~Ref.~\cite{CS2016a}). Conclusive results on the occurrence of quasipinning can be found in the truncated setting $\wedge^3[\mathcal{H}_1^{(10)}]$ and the minimal distance to the polytope boundary follows as
\begin{equation}\label{eq:Dmin1dquasi}
D_{min}(\delta,\chi) = \frac{4}{9\chi^4} \delta^2 + \frac{20}{2187}\delta^8  +\mathcal{O}(\delta^{10}).
\end{equation}
Note that due to the condition \eqref{eq:regimecondition} we have $\delta^8\gtrsim \delta^2/\chi^4$. The truncation error, given by $\sum_{k=11}^\infty\lambda_k(\delta,\chi)$, is of the order $\mathcal{O}(\delta^{10})$ which is negligible indeed.

Furthermore, taking the limit $\chi\rightarrow\infty$ as a consistency check reveals that
\begin{equation}\label{eq:limitOfDminInArbitraryDimHarmo3}
\lim_{\chi \rightarrow \infty} D_{min}(\delta,\chi) = \frac{20}{2187}\delta^8 + \mathcal{O}(\delta^{10})\,,
\end{equation}
which coincides with the quasipinning found for the strictly one-dimensional $3$-Harmonium (see Eq. (16) in Ref.~\cite{CS2013}).

The result (\ref{eq:Dmin1dquasi}) indicates that the strength of quasipinning in one-dimensional systems is only slightly reduced when adding a second spatial dimension. This is not surprising since in general adding additional spatial dimensions, whose degrees of freedom, however, are frozen, should not change physical quantities.

\section{Fully spin-polarized Harmonium in higher spatial dimensions}\label{sec:Harmonium23dNoSpin}
After having presented an analysis on quasipinning of the Harmonium in quasi-one-dimensional setups in the previous section, we are now going to consider the extension to higher dimensions. Still, we assume a complete alignment of all spins, i.e.~freezing out the spin degree of freedom.

\subsection{Two spatial dimensions and corresponding dimensional crossovers}\label{sec:2dim}
Let us start by considering the weak coupling regime. In addition we restrict this analysis to the case of an isotropic external trap, i.e.~$\omega^{(2)} =\omega^{(1)}$.
According to the definitions (\ref{eq:kappa}) and (\ref{eq:delta}), we have $\kappa\equiv \kappa^{(1)}=\kappa^{(2)}$ and $\delta\equiv \delta^{(1)}=\delta^{(2)}$, respectively. Due to the simplifications following from the duality of NONs (\ref{eq:duality}) we determine Taylor series of various NONs in the mathematically more convenient parameter $\delta$. First, we discuss in more detail the case of three fermions. The corresponding unique ground state is given by Theorem \ref{thm:gs} and its configuration follows as $\mathcal{C}=\{\{0,0\},\{1,0\},\{0,1\}\}$. The perturbational expansion of its NONs leads to the following results up to corrections of the order $\mathcal{O}(\delta^{(8)})$,
\begin{equation}
\begin{aligned}
\lambda_{1}(\delta) &= 1-\frac{32 \delta ^4}{81}+\frac{224 \delta ^6}{729}+\mathcal{O}(\delta^{8}),\\
\lambda_{2}(\delta) = \lambda_{3}(\delta) &= 1-\frac{4 \delta ^2}{9}+\frac{4 \delta ^4}{27}-\frac{152 \delta ^6}{3645}+\mathcal{O}(\delta^{8}),\\
\lambda_{4}(\delta) = \lambda_{5}(\delta)  &= \frac{4 \delta ^2}{9}-\frac{8 \delta ^4}{27}+\frac{16 \delta ^6}{135}+\mathcal{O}(\delta^{8}),\\
\lambda_{6}(\delta) = \lambda_{7}(\delta)  &= \frac{4 \delta ^4}{27}-\frac{88 \delta ^6}{729}+\mathcal{O}(\delta^{8}),\\
\lambda_{8}(\delta) = \lambda_{9}(\delta)  &= \frac{4 \delta ^4}{27}-\frac{40 \delta ^6}{243}+\mathcal{O}(\delta^{8}),\\
\lambda_{10}(\delta)  &= \frac{8 \delta ^4}{81}-\frac{16 \delta ^6}{243}+\mathcal{O}(\delta^{8}),\\
\lambda_{i\geq 11}(\delta)  &= \mathcal{O}(\delta^{6}).
\end{aligned}
\end{equation}
Note that some NONs are degenerate in pairs. This is a direct consequence of the substructure of the $1$-RDO as given by Eq.~\eqref{eq:1RDOparities} since $spec[\rho_{1,\boldsymbol{oe}}^{(f)}]=spec[\rho_{1,\boldsymbol{eo}}^{(f)}]$ in the case of an isotropic trap.
It should be also stressed that the extension of the perturbation series to deformed anisotropic traps is a bit more tedious but can be carried out in a similar fashion.

Since the spectrum $\vec{\lambda}(\delta)$ is infinite we exploit again the concept of truncation. By considering the truncated setting $\wedge^3[\mathcal{H}_1^{(10)}]$ we find a minimal distance $D_{min}(\delta)$ of $(\lambda_k(\delta))_{k=1}^{10}$ to the boundary of the corresponding polytope $\mathcal{P}_{3,10}$, given by
\begin{equation}\label{eq:D32d}
D_{min}(\delta) = \frac{8}{27}\delta^4 + \mathcal{O}(\delta^6)\,.
\end{equation}
Since the neglected NONs are of smaller order, $\mathcal{O}(\delta^6)$, this truncated (quasi)pinning analysis is conclusive: For weak coupling the ground state NONs $\vec{\lambda}(\delta)$ in the infinite-dimensional setting $(N,d)=(3,\infty)$ are not exactly on, but very close to the boundary of the allowed region $\mathcal{P}_{3,\infty}$.
This distance is indeed smaller by two orders in $\delta$ than the distance $D_{HF}$ of $\vec{\lambda}(\delta)$ to the Hartree-Fock point $\vec{\lambda}_{HF}\equiv(1,1,1,0,\ldots)$,
\begin{equation}
D_{HF}(\delta) = \frac{8}{9}\delta^2 + \frac{8}{81}\delta^4 + \mathcal{O}(\delta^6)\,.
\end{equation}
On the other hand, comparing these results on quasipinning of the $3$-Harmonium ground state in two dimensions to the remarkable result of $\delta^8$-quasipinning ($D_{min}\sim \delta^8$) found in one dimension \cite{CS2013} indicates that reduced spatial dimensionality seems to be essential for the occurrence of strong quasipinning.

In the same way as for $N=3$ particles we study ground states of the Harmonium model for $N>3$ and determine Taylor series for the corresponding NONs. We present the results in Table \ref{tab:2d} for the cases of $N=4,5,6,7$.
 \begin{table}
\arraycolsep=2.5pt\def\arraystretch{1.3}
$
\begin{array}{c|c|c|c|c|c}
{(N,n)}&\delta^{2}&\delta^{4}&\delta^{6}&D_{min}&10^{-Q}\\  \hline
(3,2)& \wedge^{2}[\mathcal{H}_1^{(4)}] & \wedge^{3}[\mathcal{H}_1^{(10)}] & \wedge^{3}[\mathcal{H}_1^{(14)}] & \propto\delta^4& \propto \delta^0 \\ \hline
(4,2)& \wedge^{3}[\mathcal{H}_1^{(7)}] & \wedge^{4}[\mathcal{H}_1^{(13)}] & \wedge^{4}[\mathcal{H}_1^{(20)}] & \propto\delta^4 & \propto \delta^2\\ \hline
(5,2)& \wedge^{3}[\mathcal{H}_1^{(7)}] & \wedge^{5}[\mathcal{H}_1^{(14)}] & \wedge^{5}[\mathcal{H}_1^{(20)}] & \propto\delta^4  & \propto \delta^0\\ \hline
(6,2)& \wedge^{3}[\mathcal{H}_1^{(7)}] & \wedge^{5}[\mathcal{H}_1^{(14)}] & \wedge^{5}[\mathcal{H}_1^{(21)}] & \propto\delta^6  & \propto \delta^0\\ \hline
(7,2)& \wedge^{4}[\mathcal{H}_1^{(9)}] & \wedge^{6}[\mathcal{H}_1^{(17)}] & \wedge^{7}[\mathcal{H}_1^{(24)}] & \propto\delta^6 & \propto \delta^0 \\
\end{array}
$
\caption{For the $N$-Harmonium ground states in $n=2$ spatial dimensions we present the `active space structures' by considering NONs with corrections to the values $1$ or $0$ up to the orders $\delta^{2}$, $\delta^{4}$ and $\delta^{6}$, respectively. The results on quasipinning are presented ($D_{min}$) as well  as its `non-triviality', quantified by the $Q$-parameter (see text).}
\label{tab:2d}
\end{table}
There, in the second, third and fourth column we present the active space structures by taking into account different orders in $\delta$.
An active space structure $\wedge^{N'}[\mathcal{H}_1^{(d')}]$ on the scale $\mathcal{O}(\delta^r)$ means that exactly $N'$ NONs, $\lambda_{N-N'+1},\ldots,\lambda_N$, and $d'-N'$ NONs, $\lambda_{N+1},\ldots,\lambda_{N+d'-N'}$, have corrections on the scales $\delta^s, s\leq r$ to the maximal value $1$ and the minimal value $0$, respectively.
It is particularly remarkable that such well-pronounced hierarchies of actives spaces exist for the Harmonium model and can even be proven analytically. In addition, it should be stressed that such hierarchies are very convenient for the (quasi)pinning analysis. For any scale of interest, we can choose the coupling such that the higher orders are sufficiently small and the corresponding truncation error becomes
negligible. In the second last column we present the strength of the quasipinning by stating the minimal distance $D_{min}$ of $\vec{\lambda}(\delta)$ to the boundary of $\mathcal{P}_{N,\infty}$, determined by exploiting the concept of truncation. For the cases $N=4,5$ we find again quasipinning of the strength $\delta^4$ which increases for $N=6,7$ to $\delta^6$-quasipinning. This increase of the quasipinning strength by adding more fermions to the trap suggests the existence of a `Pauli pressure', created by the additional particles, pressing $\vec{\lambda}$ closer to the polytope boundary. For the case of the corresponding one-dimensional system discussed in Ref.~\cite{CS2016a} the active space hierarchy was even more well-pronounced due to missing degenerate angular degrees of freedom (which typically reduce the `Pauli pressure') in agreement with the much stronger quasipinning of the order $\delta^{2N}$ for $N\geq 4$ particles found there.

As it can be inferred from the inclusion relation $\mathcal{P}\subset \Sigma$, illustrated in Fig.~\ref{fig:polytope4}, and as it has been carefully explored in Ref.~\cite{CSQ}, quasipinning by GPCs can in some cases be just a consequences of quasipinning by PEP constraints. For instance, the distance of $\vec{\lambda}$ to the polytope boundary is bounded from above by $1-\lambda_1$, the distance of $\vec{\lambda}$ to the corresponding facet $\lambda_1=1$ of the Pauli simplex $\Sigma$. Hence, it is not only important to explore and quantify quasipinning by GPCs on an absolute scale but also relative to possible quasipinning by PEP constraints. In Ref.~\cite{CSQ} a measure for such `non-triviality' of quasipinning by GPCs was constructed, the so-called $Q$-parameter: $\vec{\lambda}$ is $10^{Q(\vec{\lambda})}$ times closer to the polytope boundary than one may expect from a possibly small distance of $\vec{\lambda}$ to the boundary of the Pauli simplex $\Sigma$. The results for the $Q$-parameter are shown in the last column of Table \ref{tab:2d}. The quasipinning found for various particle numbers is in all cases, except $N=4$, `trivial'. It follows already from the approximate saturation of PEP constraints, typically $1-\lambda_1\geq 0$. Only in the case of four fermions the quasipinning by GPCs is `non-trivial', by two orders in $\delta$. Comparing this to the results for $N$ fermions in one spatial dimension \cite{CS2016a}, namely $10^{-Q(\vec{\lambda}(\delta))}\sim \delta^2$ for all $N$, suggests that the GPCs are particularly relevant in lower spatial dimensions (one dimension). In the following we find further evidence for this by comparing dimensional crossovers by ramping up one of the two trap frequencies to approach more and more the effectively one-dimensional regime.

\begin{figure}[]
\centering
\includegraphics[width=8.5cm]{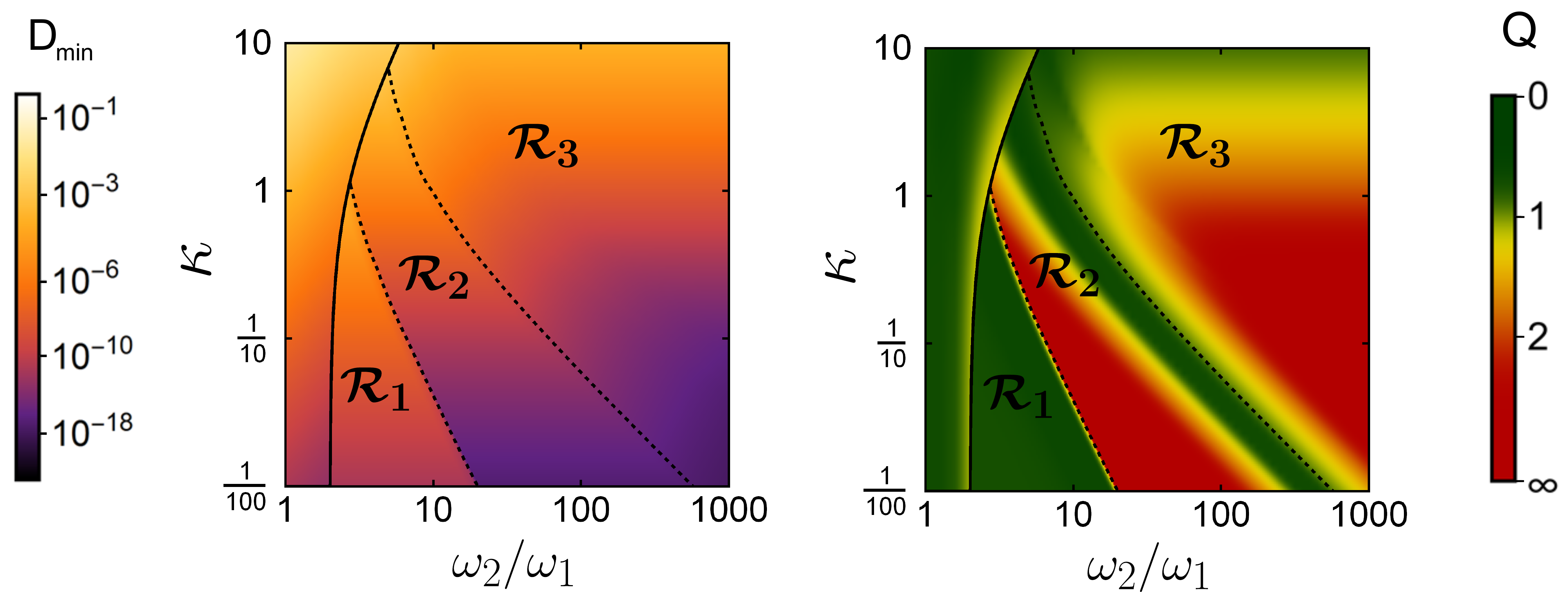}
\caption{Quasipinning results for the $3$-Harmonium ground state in two spatial dimensions with external trap frequencies $\omega_1,\omega_2$ and coupling strength $\kappa$. Minimal distance $D_{min}$ of $\vec{\lambda}$ to polytope boundary is shown on the left and its `non-triviality' is quantified on the right. The solid line represents the boundary $\kappa_{crit}(\chi)$ in between the two different ground state configurations with the effectively one-dimensional one on the right. Dashed lines mark relevant crossing of NONs (see text for more details).}
\label{fig:2to1D}
\end{figure}

We parameterize the coupling regime by $\kappa \equiv \kappa_1\geq 0$ (recall Eq.~\eqref{eq:kappa}) and the detuning $\chi=\omega_{2}/\omega_{1}\geq 1$ of the trap frequencies. We consider the case of three fermions. For smaller detunings, the ground state takes the configuration  $\boldsymbol{\mu}=\{\{0,0\},\{1,0\},\{0,1\}\}$. For larger detunings, the ground state becomes effectively one-dimensional, $\boldsymbol{\mu}'=\{\{0,0\},\{1,0\},\{2,0\}\}$ (see remarks at the end of Section \ref{sec:model}). Variation of $\chi$ while keeping $\kappa$ fixed therefore allows one to study the crossover from two to one spatial dimension.
In the $\kappa$-$\chi$ parameter plane, the  transition line between the two ground state configurations is given by
$\kappa_{crit} = -\frac{4}{3} + \frac{1}{3}\chi^2$.
Following Section \ref{sec:1RDM}, the truncated $1$-RDO has been analytically computed. For a logarithmically distributed set of data points in the $\kappa$-$\chi$ parameter plane, the NONs of truncated $1$-RDOs have then been numerically evaluated and used to determine $D_{min}(\kappa,\chi)$ and $Q(\kappa,\chi)$. The results are displayed in Figure \ref{fig:2to1D}. For all coupling parameters considered there, as well as in the figures of subsequent sections, the concept of truncation was used and the truncation errors turned out to be negligible.
The transition line between the two state configurations $\boldsymbol{\mu}$ and $\boldsymbol{\mu}'$ is marked by a solid line. In addition, there are shown two dashed lines indicating crossings of some specific (non-ordered) NONs \footnote{The left dashed line describes $\lambda_4=\lambda_5$ and the right one $\lambda_6=\lambda_7$}. In principle, there are many more crossings but according to the concept of truncation only the crossings of the largest few NONs can change the quasipinning behavior considerably.

As a first qualitative result we observe that $D_{min}$ monotonically decreases under the reduction of the fermion-fermion coupling strength $\kappa$. This is not surprising since this reduces the correlations in the system and $\vec{\lambda}$ approaches the Hartree-Fock point. For the case $\omega_1=\omega_2$ the numerical results for the whole $\kappa$-regime $[0,10]$ are well-described by the analytic perturbation-theoretical result (\ref{eq:D32d}) since $\delta=1$ corresponds to $\kappa \approx 55$. We further learn that by narrowing the trap, i.e.~by increasing $\omega_2/\omega_1$, $D_{min}$ decreases as well. This provides further evidence for the existence of a `Pauli pressure', supposed to increase whenever the number of available states around the Fermi level reduces. It should be stressed (not shown here) that even the state with configuration $\boldsymbol{\mu}=\{\{0,0\},\{1,0\},\{0,1\}\}$, being an excited state for too large detunings, begins to exhibit stronger and stronger quasipinning by increasing the detuning.

\begin{figure}[]
\centering
\includegraphics[width=7cm]{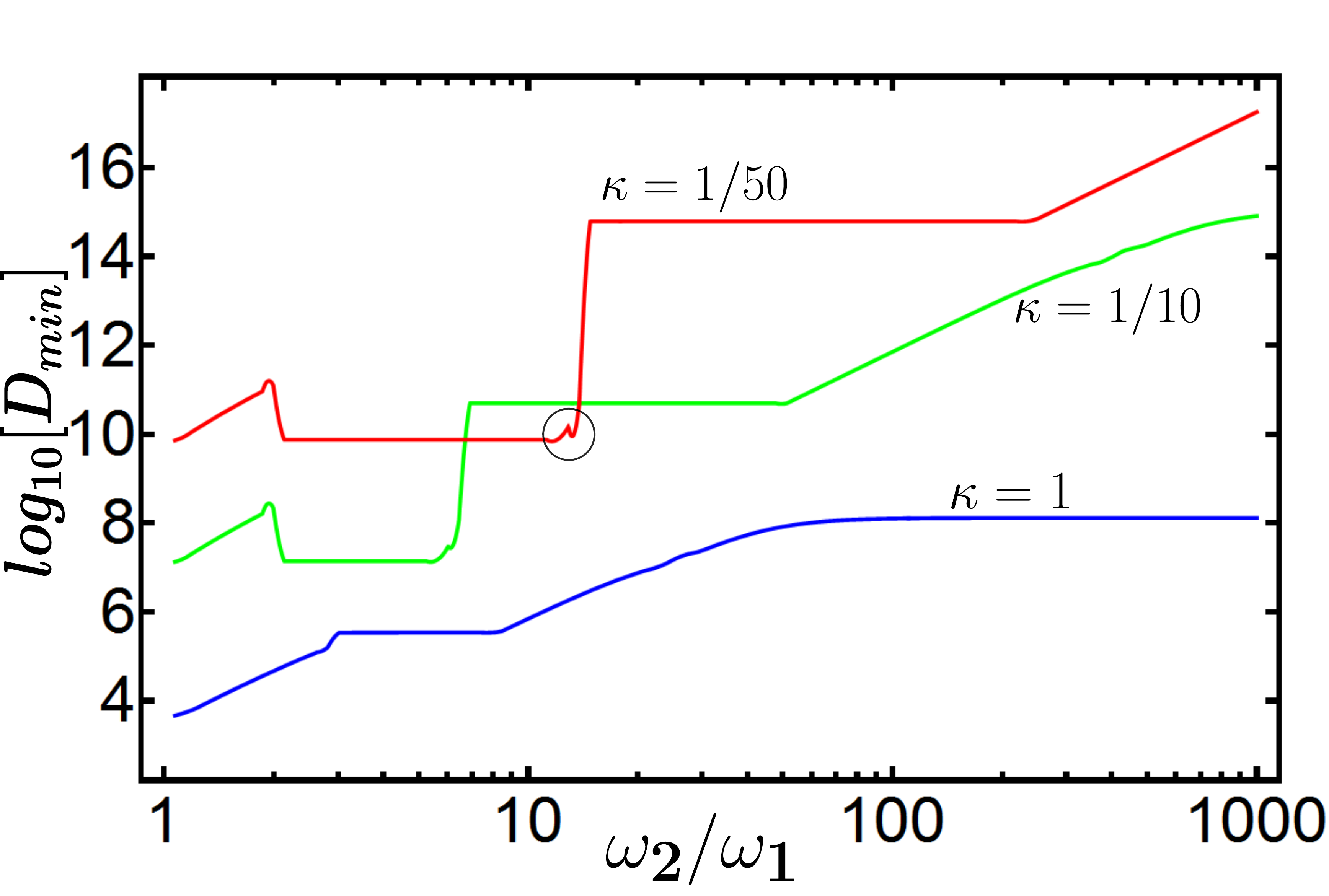}\\
\includegraphics[width=7cm]{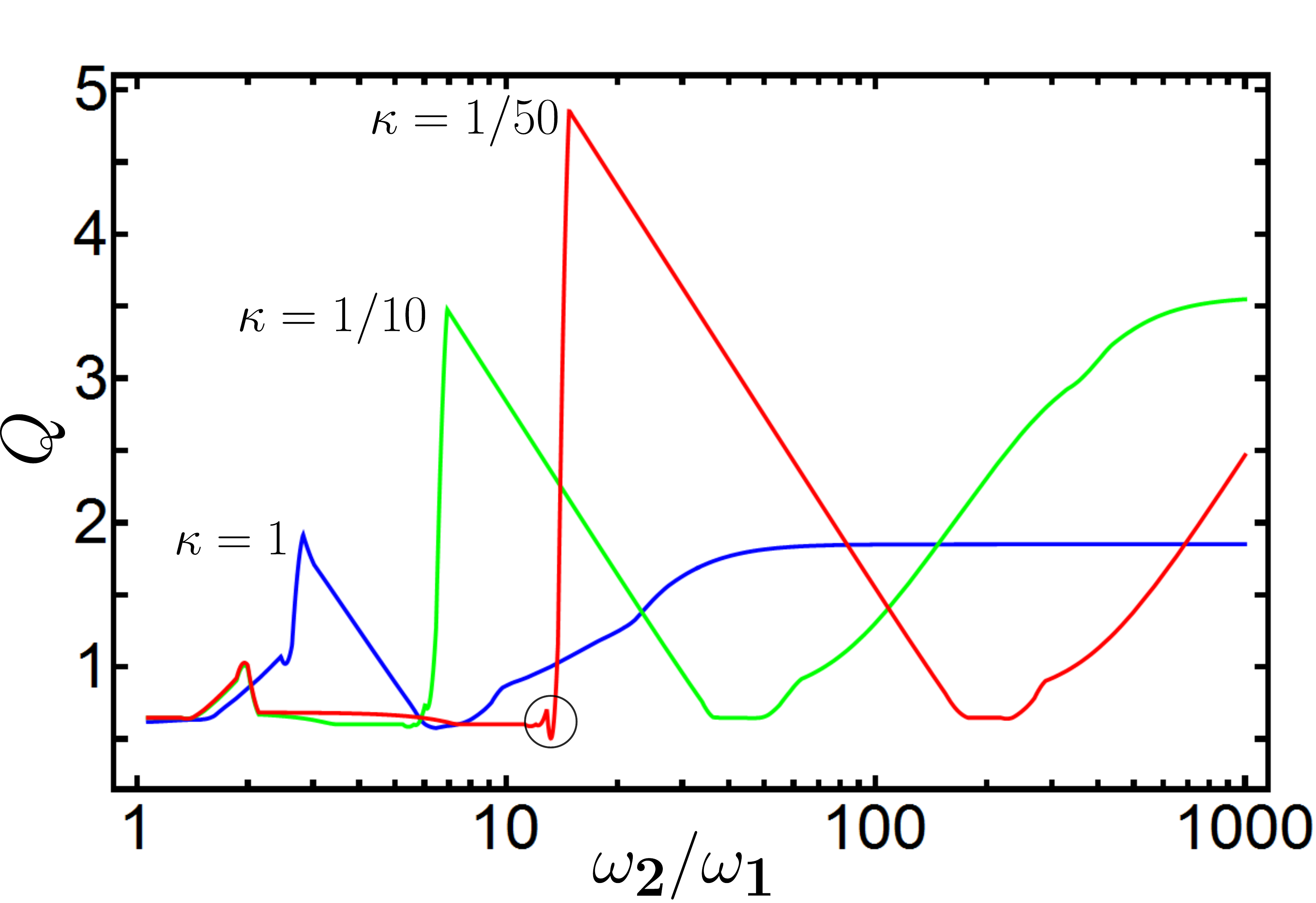}
\caption{Top: Minimal distance $D_{min}$ of $\vec{\lambda}$ to the polytope boundary as a function of the trap frequency detuning $\chi\equiv \omega_2/\omega_1$ for fixed particle-particle interaction strengths $\kappa$. Bottom: Relevance of the quasipinning by GPCs beyond Pauli's exclusion principle as measured by the $Q$-parameter.}
\label{fig:KappaSectionDmin2to1D}
\end{figure}

The relevance of the quasipinning beyond the Pauli exclusion principle as measured by the $Q$-parameter, however, shows a more complex and a different behaviour than $D_{min}$. For $\kappa \lesssim 1$ fixed, the system undergoes multiple changes from `trivial' quasipinning (green) to highly `non-trivial' quasipinning (red). Between different regimes $\mathcal{R}_i$ within the effectively one-dimensional state configuration, we observe sharp gradients in both, the $Q$-parameter as well as the minimal distance $D_{min}$. To explore those transitions further we present
fixed $\kappa$-sections in Figure \ref{fig:KappaSectionDmin2to1D}. While the change of the ground state configuration does not have a tremendous influence on the absolute quasipinning this is different for the first dashed line (in Fig.~\ref{fig:2to1D}), where $\lambda_4$ and $\lambda_5$ are coming together: Within a short $\omega_2/\omega_1$-interval, the quasipinning increases by several orders for the two exemplary cases $\kappa=1/10, 1/50$. In addition this quasipinning changes from `trivial' to highly `non-trivial'. Also crossings of smaller eigenvalues lead to non-analytic behavior of $D_{min}$ and $Q$ (see e.g.~the `black' circles) which, however, according to the concept of truncation does not change the quasipinning behavior significantly.
Finally, we conclude that for the small coupling $\kappa=1/50$ (red curve) detunings even larger than $\omega_2/\omega_1=1000$ are required to reproduce the quasipinning-results for the one-dimensional system.

The case of four particles in two spatial dimensions is contained in the next section, analyzing dimensional crossovers for $N=4$ in three spatial dimensions.

\subsection{Three spatial dimensions and corresponding dimensional crossovers}\label{sec:3dim}
The analytic results for an isotropic harmonic trap in three spatial dimensions for the regime of weak couplings $\delta$ and $N=3,4,5,6$ is presented in the Table \ref{tab:3d}.
 \begin{table}
\arraycolsep=2.5pt\def\arraystretch{1.3}
$
\begin{array}{c|c|c|c|c|c}
{(N,n)}&\delta^{2}&\delta^{4}&\delta^{6}&D_{min}&10^{-Q}\\  \hline
(3,3)& \wedge^{3}[\mathcal{H}_1^{(8)}] & \wedge^{3}[\mathcal{H}_1^{(18)}] & \wedge^{3}[\mathcal{H}_1^{(32)}] & O(\delta^4) & O(\delta^2) \\ \hline
(4,3)& \wedge^{3}[\mathcal{H}_1^{(8)}] & \wedge^{4}[\mathcal{H}_1^{(20)}] & \wedge^{4}[\mathcal{H}_1^{(34)}] & \propto\delta^4& \propto\delta^0 \\ \hline
(5,3)& \wedge^{4}[\mathcal{H}_1^{(12)}] & \wedge^{5}[\mathcal{H}_1^{(26)}] & \wedge^{5}[\mathcal{H}_1^{(45)}] & O(\delta^4)  & ? \\ \hline
(6,3)& \wedge^{5}[\mathcal{H}_1^{(14)}] & \wedge^{6}[\mathcal{H}_1^{(29)}] & \wedge^{5}[\mathcal{H}_1^{(49)}] & O(\delta^4) & ?  \\ \hline
\end{array}
$
\caption{For the $N$-Harmonium ground states in $n=3$ spatial dimensions we present the `active space structures' by considering NONs with corrections to the values $1$ or $0$ up to the orders $\delta^{2}$, $\delta^{4}$ and $\delta^{6}$, respectively. The results on quasipinning are presented ($D_{min}$) as well as its `non-triviality', quantified by the $Q$-parameter. A question mark indicates that no analysis was possible due to too large active spaces.}
\label{tab:3d}
\end{table}
In the second, third and fourth column we present the active space structures by taking into account the orders $\mathcal{O}(\delta^{2})$, $\mathcal{O}(\delta^{4})$ and $\mathcal{O}(\delta^{6})$, respectively. Comparing those active spaces with those for two spatial dimensions, shown in Tab.~\ref{tab:2d}, confirms that the additional angular degrees of freedom in $n=3$ dimensions increase significantly the dimensions of the active spaces and therefore reduce the `Pauli pressure'. This increase of the dimension reduces the chances for a \emph{conclusive} (quasi)pinning analysis. Indeed, as we can infer from the last two columns this is only possible for $N=4$. A conclusive (quasi)pinning analysis for the larger particle numbers $N=5,6$ would require the knowledge of the GPCs at least for the settings $(N,d)=(4,12)$ and $(5,14)$, respectively. For the cases $N=3,4$ we find quasipinning of the strength $\delta^4$ which might be even stronger for $N=3$ since the truncation error is of the same order, $\mathcal{O}(\delta^4)$. For $N=3$, as quantified by the $Q$-parameter, this quasipinning by GPCs is `non-trivial' by at least two orders in $\delta$. For $N=4$, however, it follows already from quasipinning by PEP constrains, $1-\lambda_1(\delta)\sim \delta^4$. For $N=5,6$ we can determine only lower bounds on the strength of the quasipinning (upper bounds on $D_{min}$): Since $1-\lambda_1\sim \delta^4$ we can conclude quasipinning by GPCs of order four or larger in $\delta$.

\begin{figure}[]
\centering
\includegraphics[width=8.0cm]{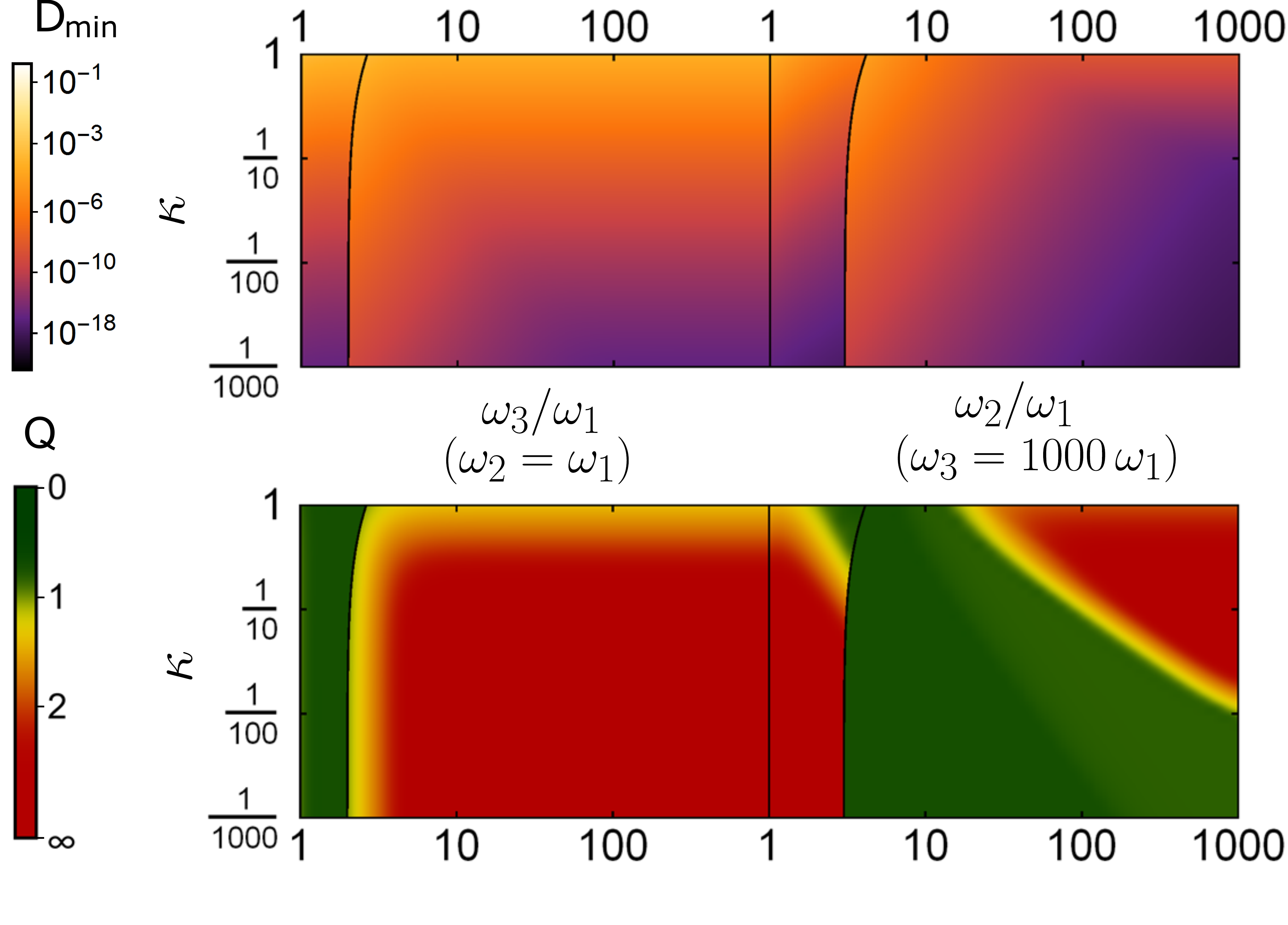}
\caption{Quasipinning results for the $4$-Harmonium ground state in three dimensions with external trap frequencies $\omega_1,\omega_2,\omega_3$ and coupling strength $\kappa$. Minimal distance $D_{min}$ of $\vec{\lambda}$ to polytope boundary is shown in the upper diagram and its `non-triviality' $Q$ is quantified in the lower diagram (see text for details). The left and right solid curve indicate a change of the ground state configuration.}
\label{fig:3to2to1D}
\end{figure}

To further explore the role of the spatial dimension for quasipinning we study dimensional crossovers for the exemplary case of four fermions. The results are shown in Fig.~\ref{fig:3to2to1D}. We first start with the isotropic trap, $\omega_1=\omega_2=\omega_3$ and ramp up continuously $\omega_3$ to the value $\omega_3=1000\,\omega_1$ while fixing $\omega_2=\omega_1$. In particular, this induces a change of the ground state configuration (first solid line) from $\boldsymbol{\mu}=\{\{0,0,0\},\{1,0,0\},\{0,1,0\},\{0,0,1\}\}$ to $\{\{0,0,0\},\{1,0,0\},\{0,1,0\},\{2,0,0\}\}$.
Then, we approach the effectively one-dimensional regime by also ramping up $\omega_2$ relative to $\omega_1$,
inducing in particular another change of the ground state configuration (last solid line) to $\{\{0,0,0\},\{1,0,0\},\{2,0,0\},\{3,0,0\}\}$. To keep the error of the truncated (quasi)pinning analysis (performed in the largest known setting $(N,d)=(4,10)$) sufficiently small we need to restrict ourselves to the coupling regime $\kappa\in [0,1]$.  Recall that in two spatial dimensions we could conclusively explore quasipinning up to couplings $\kappa=10$. This was due to the fact that the active spaces in two spatial dimensions are smaller and that the GPCs for N=3 (as analyzed in Fig.~\ref{fig:2to1D}) are already known for $d=11$.

In Fig.~\ref{fig:3to2to1D} we observe that the minimal distance $D_{min}$ of $\vec{\lambda}$ to the polytope boundary monotonically decreases with decreasing $\kappa$, essentially reflecting the  fact of reducing correlations. For every fixed state configuration $\boldsymbol{\mu}$, increasing the anisotropy of the trap increases the strength of the quasipinning as well. Intriguingly, however, at the boundaries of these regimes, the quasipinning strength drops by several orders of magnitude, which has not been the case for $N=3$ particles in two spatial dimensions (see Figure \ref{fig:2to1D}).

The most striking aspect of the (quasi)pinning analysis of this system, however, is revealed by a comparison of $D_{min}$ and $Q$ in the effectively one-dimensional configuration. While $D_{min}$ decreases with smaller $\kappa$ and larger $\omega_2/\omega_1$, the $Q$-parameter shows that the quasipinning becomes `trivial' in parameter regions with small $D_{\min}$ (c.f.~lower right corner of the diagram and that it becomes highly `non-trivial' in parameter regions with large $D_{\min}$ (c.f.~upper right corner of the diagram). This irrefutably demonstrates the importance to quantify quasipinning by GPCs \emph{beyond} quasipinning by PEP constraints.

Let us add a closing remark on degeneracies of the eigenspaces of Harmonium in higher dimensions. The 'box picture' (see Fig. \ref{fig:box}) reveals the occurrence of shell structures in Harmonium systems where an additional rotational symmetry is given whenever two or more trapping frequencies will be equal. For example $N=4$ particles in two dimensional Harmonium with $\omega_1 = \omega_2$ will have a three-fold degenerate ground state space. The obvious question whether superpositions of these states may experience a different quasipinning has been addressed by considering various superpositions, including the rotational symmetry adapted eigenstates. The quasipinning behaviour of these states was found to show the same scaling as the one for non-symmetry adapted states. They therefore have  not separately been displayed in the Tables \ref{tab:2d} and \ref{tab:3d}.

\section{Spin}\label{sec:spin}
In the previous sections and in Refs.~\cite{CS2013,CS2016a} the Harmonium model was studied for spinless/fully-polarized fermions, only, due to a good reason: By considering spinful particles with spin $S$, each spatial orbital gets a multiplicity of factor $2S+1$. This significantly reduces the `Pauli pressure' which is expected to reduces the relevance of the GPCs as well. In addition, the spin-degeneracies make an interpretation of possible quasipinning more challenging. In this section we eventually consider non-fully polarized systems. This will also allow us to induce changes of the quantum state and thus of quasipinning by ramping up an external magnetic field. In that sense, we provide new conceptual ideas for experimental realizations of the quasipinning-phenomenon.

Besides taking spin degrees into account, by extending the $N$-fermion Hilbert space to $\wedge^N[L_2[\R^n]\otimes \C^2]$ by assuming $S=1/2$, we couple the spins to an external homogenous magnetic field $\vec{B}$, described by the Zeeman term $-c/\hbar\sum_{i=1}^{N} \vec{s}_i\cdot \vec{B}$, added to the Hamiltonian (\ref{eq:ham}), where $c$ is a coupling constant \footnote{We skip the coupling of the magnetic field to the angular momentum of the fermions by assuming them to be uncharged. One could in principle extend our model to charged fermions which, however, would significantly complicate the description of the ground state for non-isotropic traps.}.

Since the Zeeman-term commutes with the remaining part of the Hamiltonian, Theorem \ref{thm:gs} and the `box-picture' can easily be extended to the spinful case:
Each orbital $\phi_{\bd{\mu}_j}^{(\bd{\tilde{\bd{l}}})}(\vec{x}_j)$ in the Slater determinant in Eq.~(\ref{eq:gs}) is multiplied by a spin state $\ket{\sigma_j}=\ket{\!\uparrow},\ket{\!\downarrow}$, defined with respect to the quantization axis $\vec{B}/B$. The box-picture is modified by considering two `box-arrays', one for $\ket{\!\uparrow}$ and one for $\ket{\!\downarrow}$. These two `box-arrays' are energetically displaced by the energy difference $g |B|$ between $\ket{\!\uparrow}$ and $\ket{\!\downarrow}$ due to the external magnetic field B. Then, the ground state configuration $\boldsymbol{\mu} = (\boldsymbol{\mu}^\uparrow,\boldsymbol{\mu}^\downarrow)$ follows again by distributing the $N$ fermions into the $N$ energetically lowest `boxes' and the ground state is given by the modified Eq.~(\ref{eq:gs}).  For instance, the ground state of $N=4$ particles in an isotropic two-dimensional trap with frequencies $\omega_1=\omega_2\equiv \omega$ with an external magnetic field $\hbar \omega \sqrt{1 +\kappa}/c<|B|<2\hbar \omega\sqrt{1 + \kappa}/c$  is described by ${\boldsymbol{\mu}^{\downarrow}=\{\{0,0\},\{1,0\},\{0,1\}\}}$ and ${\boldsymbol{\mu}^{\uparrow}=\{\{0,0\}\}}$.

\begin{figure}[]
\centering
\includegraphics[width=8.0cm]{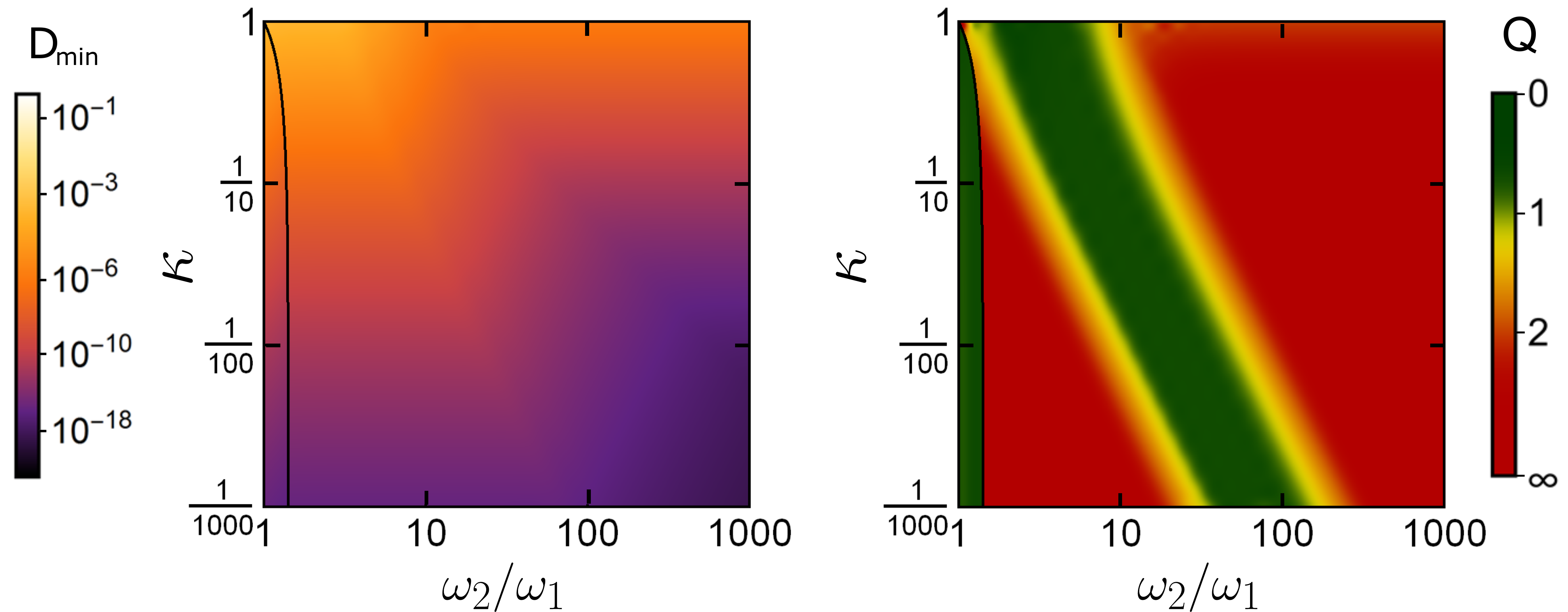}
\caption{Quasipinning results for the non-fully spin-polarized $3$-Harmonium ground state in two spatial dimensions with external trap frequencies $\omega_1,\omega_2$, coupling strength $\kappa$ and some fixed homogenous magnetic field $B$ fulfilling $\hbar \omega \sqrt{1+\kappa}< c|B| < 2 \hbar \omega \sqrt{1+\kappa}$. Minimal distance $D_{min}$ of $\vec{\lambda}$ to polytope boundary is shown in the left diagram and its `non-triviality' is quantified in the right diagram by the $Q$-parameter. The solid line marks the boundary between the effectively two- and one-dimensional ground states configuration.}
\label{fig:N3d2WithSpinDimTransition}
\end{figure}

We first investigate quasipinning for the one-dimensional case. For the regime of weak coupling $\delta$ we determine for $N=3,4$ and each spin sector the corresponding ground state. For $N=3$, depending on the field strength $B$ of the magnetic field and the coupling parameter $\delta$, the ground state can have the magnetic quantum number $M=\pm\frac{3}{2}, \pm\frac{1}{2}$ and for $N=4$ the magnetic quantum number $M=\pm2,\pm 1,0$, respectively. A thorough and conclusive (quasi)pinning analysis by using the concept of truncation shows the following universal quasipinning-behavior
\begin{equation}\label{eq:DspinM}
D_{min}(\delta) \sim \delta^{4 + 2 |M|}\,.
\end{equation}
This remarkable result for $N=3,4$ strongly supports the relevance of the proposed `Pauli pressure' for the occurrence of quasipinning: The larger the degree of polarization, $|M|$, the stronger the conflict between energy minimization and fermionic exchange symmetry and the stronger the quasipinning according to Eq.~(\ref{eq:DspinM}).

We briefly comment on the `non-triviality' of the quasipinning (\ref{eq:DspinM}) by comparing it to quasipinning by PEP constraints. For $N=3$ and both values $|M|=\frac{3}{2}, \frac{1}{2}$ the quasipinning by GPCs is `non-trivial' by two orders in $\delta$. The same holds of course for $N=4$ in case of full polarization \cite{CS2016a}. Yet, for not-fully polarized fermions, $|M|=0,1$, quasipinning by GPCs turns out to be `trivial'. In contrast to the case $M=\pm 1$, this is obvious for the case $M=0$ \footnote{For the Harmonium ground state, $M=0$ always means that every occupied `box' is doubly occupied and therefore closed. Hence, the total spin for such a configuration is $S=0$.}: It is well-known \cite{Smith66} that for spin-singlet
states the only constraints on the NONs are the PEP constraints. In other words, by restricting the $N$-fermion Hilbert space to spin-singlet states the corresponding GPCs coincide with the PEP constraints and therefore do not facilitate any `non-trivial' quasipinning by GPCs.

As an example for higher spatial dimension we consider the case of three fermions in two dimensions. We exploit the physically more relevant coupling parameter $\kappa\equiv\kappa_1$ (recall Eq.~(\ref{eq:kappa})) and the detuning $\chi \equiv \omega_2/\omega_1\geq 1$. We start with $\chi=1$ and choose the magnetic field such that $\hbar \omega \sqrt{1+\kappa}< c|B| < 2 \hbar \omega \sqrt{1+\kappa}$. This leads to a ground state with index sets $\boldsymbol{\mu}^\uparrow = \{\{0,0\},\{1,0\},\{0,1\}\}$ and $\boldsymbol{\mu}^\downarrow = \emptyset$. Increasing $\omega_2$ while keeping $\omega_1$ constant induces a dimensional crossover to an effectively one-dimensional configuration with index sets $\boldsymbol{\mu}^\uparrow = \{\{0,0\},\{0,1\}\}$ and $\boldsymbol{\mu}^\downarrow = \{\{0,0\}\}$.
The results of a (quasi)pinning analysis are shown in Figure \ref{fig:N3d2WithSpinDimTransition}. Whereas the minimal distance $D_{min}$ of $\vec{\lambda}$ to the polytope boundary becomes smaller when reducing $\kappa$ and increasing $\chi = \omega_2/\omega_1$, the behaviour of the $Q$-parameter is more complex and shows similarities to the fully spin-polarized case in Section \ref{sec:Harmonium23dNoSpin}. Again, the crossing of NONs induces different regimes leading to a high-low-high transition of the $Q$-values when increasing $\chi$ for a fixed interaction strength $\kappa$. This once more proves the significance of the $Q$-parameter as it unveils the importance of GPCs beyond the Pauli exclusion principle in parameter ranges with attributed moderate minimal distances $D_{min}$. In the regime of the two-dimensional ground state configuration, i.e.~left to the solid black line, the quasipinning is `trivial' (`green').

\section{Summary and Conclusion}\label{sec:concl}
By studying the Harmonium system, we have thoroughly explored how the spatial dimension, the particle number, the total spin and the coupling strength affect the physical relevance of the generalized Pauli constraints (GPCs).  First, in the form of Theorem \ref{thm:gs}, we succeeded in finding a compact analytical form for the ground state of this interacting $N$ fermion system. This then allowed us to determine the natural occupation numbers analytically by perturbation theory for the regime of small coupling strengths $\delta$ and by an exact numerical approach for medium and strong coupling strengths.

Given the set $\vec{\lambda}\equiv(\lambda_k)_{k=1}^\infty$ of NONs for the ground state of some Harmonium system and a fixed coupling we have explored whether $\vec{\lambda}$ (approximately) saturates some of the GPCs. Since such (quasi)pinning, as quantified by the minimal distance $D_{min}$ of $\vec{\lambda}$ to the boundary of the allowed region (polytope $\mathcal{P}$), has remarkable physical consequences, this would then confirm the physical relevance of the GPCs. Since the GPCs are known so far only for $1$-particle Hilbert spaces of dimension $d<11$ we exploited the concept of truncation: All occupation numbers sufficiently close to $1$ or $0$ can be neglected and possible (quasi)pinning is then explored in the truncated setting. It turned out that for most Harmonium systems and not too strong couplings the corresponding active spaces are sufficiently low-dimensional to facilitate a conclusive analysis of (quasi)pinning. In addition, since the GPCs imply the Pauli exclusion principle (PEP) constraints (whose relevance is already well-known) we quantify the quasipinning by GPCs \emph{beyond} quasipinning by PEP constraints by using the $Q$-parameter \cite{CSQ}.

In general, for the regime of small coupling strengths $\delta \ll 1$ we found that the active spaces in two and three spatial dimensions are significantly larger than for the one-dimensional case studied in Ref.~\cite{CS2016a}. This is due to the additional degenerate orbital degrees of freedom. Moreover, similar to the one-dimensional case, there are well-pronounced hierarchies of active spaces defined by considering corrections of different orders $\mathcal{O}(\delta^r)$ of the NONs to $1$ and $0$, respectively. These specific shell-structures make the concept of truncation even more powerful for the Harmonium systems.

For the case of $n=2$ spatial dimensions and spinless fermions we found for $N=3,4,5$ fermions quasipinning of strength $D_{min}\sim \delta^4$ which increased to $\delta^6$-quasipinning for $N=6,7$. Comparing this quasipinning by GPCs to quasipinning of the less-restrictive PEP constraints shows that our findings are only `non-trivial' for $N=4$, namely by two orders in $\delta$. For the case of three spatial dimensions the active space dimensions increase even further which does not allow us anymore to perform a conclusive quasipinning analysis for several $N$. For $N=3,4$ we find again $\delta^4$-quasipinning which might, at least in principle, be even stronger for $N=3$ since the corresponding truncation error is of the same order, $\mathcal{O}(\delta^4)$. The quasipinning turns out to be `trivial' for $N=4$ but  `non-trivial' by at least two orders in $\delta$ for $N=3$.

It is also instructive to compare those findings for two and three spatial dimensions to those in one dimension, showing quasipinning described by $D_{min}\sim \delta^8$ for $N=3$ \cite{CS2013} and $D_{min}\sim \delta^{2N}$ for $N\geq 4$ \cite{CS2016a}.
The increase of quasipinning by reducing the spatial dimension or by adding more fermions to the trap suggests that quasipinning emerges from a `Pauli pressure'. Such `Pauli pressure', which still needs to be formally defined and carefully worked out,
describes for ground states the conflict of energy minimization and fermionic exchange symmetry (antisymmetry) from the $1$-particle picture's viewpoint. Indeed, this conflict reduces by increasing the spatial dimension (leading to additional degenerate orbital degrees of freedom) and increases by adding more fermions to the trap.

By detuning the trap frequencies we explored crossovers between Harmonium systems of different effective spatial dimensions allowing us to also analyze quasipinning for `intermediate' dimensions. From a qualitative viewpoint, we found further evidence for the `Pauli pressure' being the origin of quasipinning: The stronger the detuning between the trap frequencies and the smaller the fermion-fermion coupling, the stronger the quasipinning. While the absolute quasipinning, as measured by the minimal distance $D_{min}$ of $\vec{\lambda}$ to the polytope boundary behaves mainly monotonically as function of the detunings and the coupling strength, the $Q$-parameter shows a much more complex behaviour: While keeping the coupling strength constant, ramping up ´the detuning can lead to multiple crossovers between `non-trivial' and `trivial' quasipinning. Although such less monotone behavior seems to be more difficult to understand it clearly shows the importance of the $Q$-parameter for a genuine quantification of the relevance of GPCs in concrete systems.

In the final section, Sec.~\ref{sec:spin}, we eventually included the spin-degree of freedom as well. By varying an external magnetic field coupling to the spins of the uncharged fermions we can induce transitions between states of different spin polarization. The main result, for not to strong couplings $\delta$ and $N=3,4$ states $D_{min}\sim \delta^{4 +2 |M|}$ (Eq.~(\ref{eq:DspinM})): The larger the total magnetization ($M$) the stronger the quasipinning. This remarkable universal relation confirms again the role of the `Pauli pressure' for quasipinning since increasing the degree of polarization reduces the effective number of available states around the Fermi level.

The findings on the spinful fermions provide a general idea for an experimental realization and verification of (quasi)pinning. In a first step, a Harmonium-like system of spinful fermions shall be prepared in its ground state exhibiting quasipinning. Then, by coupling this system to an external oscillating magnetic field the system's ground state and its corresponding vector $\vec{\lambda}$ of NONs are perturbed. Due to the strong quasipinning of $\vec{\lambda}$ to the polytope boundary this perturbation $\Delta \vec{\lambda}$ can be directed for any choice of the perturbation (here e.g.~the field-polarization) only parallel but not perpendicular to the polytope boundary (see also Ref.~\cite{CS2015Hubbard}). Without the knowledge of the GPCs and the corresponding polytope $\mathcal{P}$ such behaviour of $\vec{\lambda}$ restricted to a hyperplane looks magical. This is conceptually very similar to the prohibited decay of valence electrons to lower lying energy shells in atoms due to the more elementary Pauli exclusion principle.

\begin{acknowledgements}
We would like to thank C.~Benavides-Riveros and J.~Klassen for helpful discussions.
We gratefully acknowledge financial support from
the Friedrich-Naumann-Stiftung and Christ Church Oxford (FT), the Oxford Martin School, the NRF (Singapore), the MoE (Singapore) and the EU Collaborative Project TherMiQ (Grant Agreement 618074)
(VV), the Swiss National Science Foundation (Grant P2EZP2 152190) and the Oxford Martin Programme on Bio-Inspired Quantum Technologies (CS).
\end{acknowledgements}

\bibliography{bibliography,comments}
\onecolumngrid

\appendix
\section{Derivation of the Harmonium ground state}\label{app:gs}
In this appendix we provide an elegant proof of Theorem \ref{thm:gs}, i.e.~we derive the explicit form of the ground state(s) of Harmonium~\eqref{eq:ham}.

First, we consider the Hamiltonian (\ref{eq:ham}) on the $N$-particle Hilbert space $\mathcal{H}_N\equiv {L_2[\R^n]}^{\otimes^N}$, i.e.~without any exchange symmetry. On that space, it can easily be diagonalized by decoupling the harmonic oscillators by introducing center of mass ($y_1^{(\alpha)}$) and `relative' coordinates ($y_k^{(\alpha)}, k=2,\ldots,N$) for all $n$ spatial dimensions $\alpha=1,\ldots,n$ (see e.g.~Ref.~\cite{CS2013NO} for the case $n=1$). The corresponding eigenstates follow as
\begin{equation}\label{eq:states}
\Phi_{\bd{\nu}}^{(\bd{\omega},\bd{K})}(\vec{x}_1,\ldots,\vec{x}_N) = \mathcal{N}_{\bd{\nu}}^{(\bd{\omega},\bd{K})}\,\cdot\left[\prod_{\alpha=1}^{n}\varphi_{\nu_1^{(\alpha)}}^{(l^{(\alpha)})}( y_1^{(\alpha)}(x_1^{(\alpha)},\ldots,x_N^{(\alpha)}))\right]
\cdot \left[
 \prod_{\alpha=1}^{n}\prod_{k=2}^{N} \varphi_{\nu_{k}^{(\alpha)}}^{(\tilde{l}^{(\alpha)})}( y_k^{(\alpha)}(x_1^{(\alpha)},\ldots,x_N^{(\alpha)})) \right]\,,
\end{equation}
with $\bd{\nu}\equiv(\nu_k^{(\alpha)})$, $\nu_k^{(\alpha)} \in \N_0$, $k=1,\ldots,N$, $\alpha=1,\ldots,n$, $\bd{\omega}\equiv (\omega^{\alpha})$, $\bd{K}\equiv (K^{(\alpha)})$ and $\mathcal{N}_{\bd{\nu}}^{(\bd{\omega},\bd{K})}$ a normalization constant. Since the dependence of $\Phi_{\bd{\nu}}^{(\bd{\omega},\bd{K})}$ on the couplings $(\bd{\omega},\bd{K})$ becomes relevant below we made it explicit.
We also introduced for each spatial dimension $\alpha$ the corresponding length $l^{(\alpha)} = \sqrt{\frac{\hbar}{m \omega^{(\alpha)}}}$ for the center of mass and the lengths $\tilde{l}^{(\alpha)} = \sqrt{\frac{\hbar}{m \tilde{\omega}^{(\alpha)}}}$ for the relative motion between the particles. The corresponding frequencies are given by
\begin{equation}\label{eq:omegar}
  \tilde{\omega}^{(\alpha)} \equiv \sqrt{(\omega^{(\alpha)})^2 +\frac{N K^{(\alpha)}}{m}}\,.
\end{equation}

For the following, it will be crucial that the center of mass coordinates $y_1^{(\alpha)}(x_1^{(\alpha)},\ldots,x_N^{(\alpha)})$ are symmetric in the physical coordinates $x_1^{(\alpha)},\ldots,x_N^{(\alpha)}$ and the precise form of the `relative' coordinate functions $y_k^{(\alpha)}, k=2,\ldots,N$ will not be relevant. The corresponding energy of the eigenstates \eqref{eq:states} are given (up to a constant energy shift) by
\begin{equation}\label{eq:energies}
E_{\bd{\nu}}^{(\bd{\omega},\bd{K})} = \sum_{\alpha=1}^{n} \hbar\omega^{(\alpha)}\nu_1^{(\alpha)} + \sum_{\alpha=1}^{n} \hbar \tilde{\omega}^{(\alpha)}\,\sum_{k=2}^{N}\nu_{k}^{(\alpha)}\,.
\end{equation}
By introducing the corresponding coupling parameter space
\begin{equation}\label{eq:couplspace}
  \Omega \equiv \{(\bd{\omega},\bd{K})\in (\R^+)^n\!\times \R^n\,|\, \forall \alpha:\, N K^{(\alpha)}> -m (\omega^{(\alpha)})^2\}\,,
\end{equation}
we observe that most of the energy branches $E_{\bd{\nu}}^{(\bd{\omega},\bd{K})}$ are degenerate on $\Omega$. Accordingly, we introduce equivalence classes $[\bd{\nu}]$ by identifying the index sets $\bd{\nu}$ and $\bd{\nu}'$ of quantum numbers whenever their energies values \eqref{eq:energies} are identical on $\Omega$. Moreover, we introduce the corresponding eigenspace,
\begin{equation}\label{eq:eigenspaces}
  \mathcal{H}_{[\bd{\nu}]}^{(\bd{\omega},\bd{K})} \equiv \mbox{span}\left(\Big\{\Phi_{\bd{\nu}'}^{(\bd{\omega},\bd{K})}\,|\,\bd{\nu}'\in [\bd{\nu}]\Big\}\right)\,,
\end{equation}
which are all finite-dimensional and depend analytically on $(\bd{\omega},\bd{K}) \in \Omega$.

In order to find the fermionic ground state for arbitrary but fixed $(\bd{\omega},\bd{K}) \in \Omega$ one may consider the projection of the set of $N$-particle eigenstates \eqref{eq:states} onto the fermionic subspace
\begin{equation}\label{eq:spaceemb}
  \mathcal{H}_N^{(f)} \equiv \wedge^N [L_2[\R^n]] \lneq \mathcal{H}_N\equiv L_2[\R^n]^{\otimes^N}\,.
\end{equation}
More precisely, one needs to determine the index set $\bd{\nu}$ which minimizes the energy function \eqref{eq:energies} while still having non-vanishing support on $\mathcal{H}_N^{(f)}$, i.e.~$\mathcal{A}_{N} \mathcal{H}_{[\bd{\nu}]}^{(\bd{\omega},\bd{K})}\neq 0$.

Due to the non-trivial dependency of the `relative' coordinates $y_{i}^{(\alpha)}$ on the physical coordinates $x_{j}^{(\alpha)}$ it proves to be challenging to simplify the resulting expression of the action of the antisymmetrisation operator $\mathcal{A}_N$ on the $N$-particle eigenstates of Eq.~\eqref{eq:states}.
In the following we present an elegant, systematic way for determining the fermionic ground state:
\begin{enumerate}
  \item Due to the specific structure \eqref{eq:states}, separating symmetric center of mass coordinates from `relative' coordinates, the fermionic ground state lies in an eigenspace $ \mathcal{H}_{[(\vec{\nu}_1,\ldots,\vec{\nu}_N)]}^{(\bd{\omega},\bd{K})}$ with zero center of mass excitations, i.e.~$\vec{\nu}_1=\vec{0}$.
  \item The main idea is now to relate the fixed coupling $(\bd{\omega},\bd{K})$ to another one with zero interaction $(\bd{\omega}',\bd{K}')\equiv (\bd{\omega}',\bd{0})$ such that
\begin{equation}\label{eq:couplr}
  \omega'^{(\alpha)}\equiv \sqrt{(\omega^{(\alpha)})^2+\frac{N K^{(\alpha)}}{m}}\,,\quad\alpha=1,\ldots,n\,.
\end{equation}
The coupling parameters $(\bd{\omega}',\bd{0})$ are chosen in such a way that they lead to the same frequencies for the relative motion as $(\bd{\omega},\bd{K})$. Indeed, since $K'^{(\alpha)}=0$ we find
\begin{equation}\label{eq:omegarel0}
  \tilde{\omega}'^{(\alpha)} = \sqrt{(\omega'^{\alpha})^2+\frac{N K'^{(\alpha)}}{m}} = \sqrt{(\omega^{(\alpha)})^2+\frac{N K^{(\alpha)}}{m}} = \tilde{\omega}^{(\alpha)}\,, \quad\alpha=1,\ldots,n\,.
\end{equation}
This then implies for the energies
\begin{equation}\label{eq:energiesrel}
  E_{(\vec{0},\vec{\nu}_2,\ldots,\vec{\nu}_N)}^{(\bd{\omega},\bd{K})} = E_{(\vec{0},\vec{\nu}_2,\ldots,\vec{\nu}_N)}^{(\bd{\omega}',\bd{0})}\,,\quad \forall \vec{\nu}_2,\ldots,\vec{\nu}_N\,.
\end{equation}
Moreover, due to the specific structure of the eigenstates (\ref{eq:states}) we can relate the corresponding eigenspaces as well,
\begin{equation}\label{eq:spacesrel}
   \mathcal{H}_{[(\vec{0},\vec{\nu}_2,\ldots,\vec{\nu}_N)]}^{(\bd{\omega},\bd{K})} =  G^{(\bd{\omega},\bd{K})} \mathcal{H}_{[(\vec{0},\vec{\nu}_2,\ldots,\vec{\nu}_N)]}^{(\bd{\omega}',\bd{0})}\,,
\end{equation}
where $G^{(\bd{\omega},\bd{K})}$ describes the multiplication by the Gaussian factor $\exp\left[{-\frac{N}{2} \sum_{\alpha=1}^{n} \left(\frac{1}{(l'^{(\alpha)})^2}-\frac{1}{(l^{(\alpha)})^2}\right) (X^{(\alpha)})^2}\right]$ (see Ref.~\cite{CS2013NO})
and $X^{(\alpha)} \equiv y_1^{(\alpha)}\equiv \frac{1}{N}\,(x_1^{(\alpha)}+\ldots+x_N^{(\alpha)})$ is the center of mass coordinate in $\alpha$-direction.
  \item Since $(\bd{\omega}',\bd{K}')=(\bd{\tilde{\omega}},\bd{0})$ describes non-interacting fermions we can easily determine the fermionic ground state for this case. The corresponding Schr\"odinger equation turns into an effectively $1$-fermion equation describing a single harmonic oscillator in $n$-spatial dimensions with frequencies $\tilde{\omega}^{(\alpha)}$ and with $n$-dimensional Hermite functions $\phi_{\bd{\mu}}^{(\bd{\tilde{l}})}(\vec{x})\equiv \prod_{\alpha=1}^n \varphi_{\mu^{(\alpha)}}^{(\tilde{l}^{(\alpha)})}(x^{(\alpha)})$ as eigenstates with corresponding energy $\varepsilon_{\bd\mu}=\sum_{\alpha=1}^n(\mu^{(\alpha)}+\frac{1}{2})\hbar\tilde{\omega}^{(\alpha)}$.
The corresponding fermionic ground state for $(\bd{\tilde{\omega}},\bd{0})$ is then given by the `configuration state', i.e.~by the Slater determinant obtained by distributing the $N$ particles in $N$ energetically lowest states $\phi_{\bd{\mu}}^{(\bd{\tilde{l}})}$. By denoting the corresponding sets of quantum numbers by $\bd{\mu}_1=\bd{0},\ldots,\bd{\mu}_N$, the ground state reads
\begin{equation}\label{eq:gs0}
  \Psi_{gs}^{(\bd{\tilde{\omega}},\bd{0})} = \phi_{\bd{\mu}_1}^{(\bd{\tilde{l}})}\wedge \ldots \wedge \phi_{\bd{\mu}_N}^{(\bd{\tilde{l}})}\,.
\end{equation}
It should be also stressed that for generic couplings $(\bd{\omega},\bd{K})\in \Omega$ the ground state for $(\bd{\tilde{\omega}},\bd{0})$ is unique because the frequencies $\tilde{\omega}^{(\alpha)}$ are generically incommensurate.
\item Eq.~\eqref{eq:gs0} allows us to find the ground state for $(\bd{\omega},\bd{K})$ as well. This is based on the previous points, essentially Eqs.~\eqref{eq:energiesrel}, \eqref{eq:spacesrel}, implying \begin{equation}\label{eq:gsspacerel}
      \Psi_{gs}^{(\bd{\tilde{\omega}},\bd{0})} \in \mathcal{H}_{[(\vec{0},\vec{\nu}_2,\ldots,\vec{\nu}_N)]}^{(\bd{\tilde{\omega}},\bd{0})} \quad\Leftrightarrow \quad\Psi_{gs}^{(\bd{\omega},\bd{K})} \in \mathcal{H}_{[(\vec{0},\vec{\nu}_2,\ldots,\vec{\nu}_N)]}^{(\bd{\omega},\bd{K})}\,.
    \end{equation}
    Hence,
    \begin{equation}\label{eq:gsrel}
       \mathcal{A}_N  \mathcal{H}_{[(\vec{0},\vec{\nu}_2,\ldots,\vec{\nu}_N)]}^{(\bd{\omega},\bd{K})}
      = \mathcal{A}_N G^{(\bd{\omega},\bd{K})} \mathcal{H}_{[(\vec{0},\vec{\nu}_2,\ldots,\vec{\nu}_N)]}^{(\bd{\tilde{\omega}},\bd{0})}
      = G^{(\bd{\omega},\bd{K})} \mathcal{A}_N \mathcal{H}_{[(\vec{0},\vec{\nu}_2,\ldots,\vec{\nu}_N)]}^{(\bd{\tilde{\omega}},\bd{0})}
    \end{equation}
    and therefore
    \begin{equation}\label{eq:gsrelfinal}
      \Psi_{gs}^{(\bd{\omega},\bd{K})} = G^{(\bd{\omega},\bd{K})} \Psi_{gs}^{(\bd{\tilde{\omega}},\bd{0})}\,,
    \end{equation}
    which is nothing else than Eq.~(\ref{eq:gs}) in Theorem \ref{thm:gs}. It should be also stressed that the same conclusions also hold in case of degenerate fermionic ground states. Notice also, that Theorem \ref{thm:gs} and Eq.~(\ref{eq:gs}) do not hold for the excited fermionic states. The subtle difference between the ground state and excited states is that in case of ground states we can assume zero center of mass excitations, $\vec{\nu}=0$ in contrast to excited states. In the latter case, by following the same derivation the energy eigenspace for couplings $(\bd{\tilde{\omega}},\bd{0})$ has additional degeneracies coming from center of mass excitations.
\end{enumerate}

\section{Derivation of the fermionic $1$-RDO of $N$-Harmonium in $n$ spatial dimensions}\label{app:1RDO}
Starting with the general form of the ground state wave function of $N$-Harmonium in $n$ spatial dimensions, Eq.~\eqref{eq:gs}, the partial trace over $(N-1)$ fermions of the $N$-particle density operator $\rho_N = \Psi^\ast\Psi$ will be computed by integrating over the particle positions $\vec{x}_i\in\R^n$ with labels $i = 2,\ldots,N$. Using the Hubbard-Stratonovich relation for each spatial dimension $\gamma=1,\ldots,n$,
\begin{equation}
\exp[a^{(\gamma)}(x_1^{(\gamma)}  + \ldots + x_N^{(\gamma)})^2] = \sqrt{\frac{a^{(\gamma)}}{\pi}} \int{dy^{(\gamma)} \exp[-a^{(\gamma)} (y^{(\gamma)})^2 + 2 a^{(\gamma)} y^{(\gamma)} (x_1^{(\gamma)}  + \ldots + x_N^{(\gamma)})]},
\end{equation}
yields (see Ref.~\cite{CS2013NO})
\begin{align}\label{eq:FullFermionic1RDO}
\rho^{(f)}_1&\left(\vec{x};\vec{x}'\right)= |\mathcal{N}|^2 \cdot \underbrace{\prod_{\gamma=1}^{d}{\left(\exp \left[(\frac{1}{N^2}B^{(\gamma)}+C^{(\gamma)}-A^{(\gamma)})\left((x^{(\gamma)})^2 + (x'^{(\gamma)})^2\right) + 2C^{(\gamma)}_{N}x^{(\gamma)} x'^{(\gamma)}\right]\right)}}_{\rho_{1}^{(b)}\left(\vec{x};\vec{x}'\right) }\\
&\underbrace{\sum_{i=1}^{N}{}\left( \prod_{\gamma = 1}^{d}{}\int{}d u_{\gamma}\left\{ e^{- u_\gamma^2}
\frac{H_{\mu_{i}^{(\gamma)}}(p_\gamma u_{\gamma} + r_\gamma(x^{(\gamma)},x'^{(\gamma)}))H_{\mu_i^{(\gamma)}}(p_\gamma u_{\gamma} + r_\gamma(x'^{(\gamma)},x^{(\gamma)}))}{2^{\mu_i^{(\gamma)}} ({\mu_i^{(\gamma)}})!}\right\} \right)}_{\equiv F\left(\vec{x};\vec{x}'\right) } \nonumber,
\end{align}
In this expression, $H_{\mu_i^{(\gamma)}}$ denotes the Hermite polynomial of degree $\mu_i^{(\gamma)}$, where $\{\bd{\mu}_i\}_{i=1}^N$ is the set of quantum number vectors introduced in Section \ref{sec:model}. The quantities $A^{(\gamma)}, B^{(\gamma)}, C^{(\gamma)}$, $p$ and $r(\cdot, \cdot)$ are given by
\begin{align}
A^{(\gamma)}=& \: \frac{1}{2 (\tilde{l}^{(\gamma)})^2},\\
B^{(\gamma)} =& \:  \frac{N}{2}\left(\frac{1}{(\tilde{l}^{(\gamma)})^2}
-\frac{1}{(l^{(\gamma)})^2}\right),\\
C^{(\gamma)} =& \:\frac{(N-1){(B^{(\gamma)})}^2 }{2N^2(N^2A^{(\gamma)}-(N-1)B^{(\gamma)})},\\
p_\gamma =& \:\sqrt{\frac{B^{(\gamma)}}{N^2A^{(\gamma)}-B^{(\gamma)} (N-1)}}\quad \text{and} \\
r_\gamma(v,w) =& \:\sqrt{2A^{(\gamma)}}\left(v- \frac{B^{(\gamma)}}{2(N^2A^{(\gamma)}-(N-1)B^{(\gamma)})}(v+w)\right)\,,
\end{align}
where we suppressed the index $N$ of $A^{(\gamma)}, B^{(\gamma)}$ and $C^{(\gamma)}$. $l^{(\gamma)}$ and $\tilde{l}^{(\gamma)}$
denote again the length scale in $\gamma$-direction for the center of mass and the relative motion, respectively.

The $1$-RDO associated with the ground state of Harmonium for spinful fermions can be derived by tracing out spin and spatial degrees of freedom by applying the same ideas to the spatial part as in the case of fully spin-polarized Harmonium Eq.~\eqref{eq:FullFermionic1RDO}:

\begin{align}\label{oneRDOFourthExpression}
\rho_1^{(f)}&\left(\vec{x}, \sigma;\vec{x}', \sigma' \right) = |\mathcal{N}|^2 \cdot \underbrace{\prod_{\gamma=1}^{d}{\left(\exp \left[(\frac{1}{N^2}B^{(\gamma)}+C^{(\gamma)}-A^{(\gamma)})\left((x^{(\gamma)})^2 + (x'^{(\gamma)})^2\right) + 2C^{(\gamma)}x^{(\gamma)} x'^{(\gamma)}\right]\right)}}_{\rho_{1}^{(b)}\left(\vec{x};\vec{x}'\right) }\\
&\left[\underbrace{\sum_{i=1}^{N^{\uparrow}}{}\left( \prod_{\gamma = 1}^{d}{}\int{}d u_{\gamma}\left\{ e^{- u_\gamma^2}
\frac{H_{{\mu_{i}^\uparrow}^{(\gamma)}}(p_\gamma u_{\gamma} + r_\gamma(x^{(\gamma)}, x'^{(\gamma)}))H_{{\mu_{i}^\uparrow}^{(\gamma)}}(p_\gamma u_{\gamma} + r_\gamma(x'^{(\gamma)},x^{(\gamma)}))}{2^{{\mu_{i}^\uparrow}^{(\gamma)}} ({{\mu_{i}^\uparrow}^{(\gamma)}})!}\right\} \right)}_{\equiv F^{\uparrow}\left(\vec{x};\vec{x}'\right)} \ket{\!\uparrow}\bra{\uparrow\!} \right.\nonumber\\
& \left. + \: \underbrace{\sum_{i=1}^{N^{\downarrow}}{}\left( \prod_{\gamma = 1}^{d}{}\int{}d u_{\gamma}\left\{ e^{- u_\gamma^2}
\frac{H_{{\mu_{i}^\downarrow}^{(\gamma)}}(p_\gamma u_{\gamma} + r_\gamma(x^{(\gamma)},x^{(\gamma)}))H_{{\mu_{i}^\downarrow}^{(\gamma)}}(p_\gamma u_{\gamma} + r_\gamma(x'^{(\gamma)},x^{(\gamma)}))}{2^{{\mu_{i}^\downarrow}^{(\gamma)}} ({{\mu_{i}^\downarrow}^{(\gamma)}})!}\right\} \right)}_{\equiv F^{\downarrow}\left(\vec{x};\vec{x}'\right)}\ket{\!\downarrow}\bra{\downarrow\!}\right].\nonumber
\end{align}

In this expression, the $\boldsymbol{\mu}^{\uparrow}_i$ and $\boldsymbol{\mu}^{\downarrow}_j$ quantum number vectors represent the set of particles with spins parallel and antiparallel to the external magnetic field.

\section{Natural occupation numbers for the $3$-Harmonium ground state in one dimension}\label{app:NONsPRL}
For the sake of self-containedness  of this work we recall the natural occupation numbers for the $3$-Harmonium ground state
presented in Ref.~\cite{CS2013}. They are given by the following series up to corrections of the order $\mathcal{O}(\delta^{10})$, where the coupling parameter $\delta$ is defined in Eq.~(\ref{eq:delta}),
\begin{eqnarray}
1-\lambda_1 &= & \frac{40}{729} {\delta}^6 - \frac{1390}{59049} {\delta}^8 + \mathcal{O}(\delta^{10}) \nonumber \\
1-\lambda_2 &= & \frac{2}{9} {\delta}^4 - \frac{232}{729}{\delta}^6 + \frac{3926}{10935} {\delta}^8 +\mathcal{O}(\delta^{10}) \nonumber  \\
1-\lambda_3 &=& \frac{2}{9}{\delta}^4 - \frac{64}{243}{\delta}^6 + \frac{81902}{295245}{\delta}^8 +\mathcal{O}(\delta^{10}) \nonumber \\
\lambda_4 &= & \frac{2}{9}{\delta}^4 - \frac{64}{243}{\delta}^6 + \frac{73802}{295245}{\delta}^8 + \mathcal{O}(\delta^{10}) \nonumber \\
\lambda_5 &= &\frac{2}{9} {\delta}^4 - \frac{232}{729} {\delta}^6 + \frac{3976}{10935} {\delta}^8 +\mathcal{O}(\delta^{10}) \nonumber \\
\lambda_6 &= & \frac{40}{729} {\delta}^6 - \frac{2200}{59049} {\delta}^8 + \mathcal{O}(\delta^{10}) \nonumber \\
\lambda_7 &= & \frac{80}{2187} {\delta}^8 + \mathcal{O}(\delta^{10}) \nonumber \\
\lambda_k &= & \mathcal{O}(\delta^{2k-6})\,,\quad \mbox{for}\,k\geq 8\,. \label{eq:NONsPRL}
\end{eqnarray}

\end{document}